\documentstyle[epsfig,referee]{l-aa}

\begin{document}

   \thesaurus{03         
              (11.09.1 NGC 4321, NGC 4254;  
               11.19.2;  
               11.19.5;
               11.19.6;
               08.06.2)  
                } 

   \title{Star formation rates, efficiencies and initial mass functions in spiral galaxies}

  \subtitle{I. Method}

   \author{Jordi Cepa \inst{1,2} and Fernando de Pablos \inst{1} 
          }

   \offprints{Fernando de Pablos}

   \institute{$^1$Instituto de Astrof\'{\i}sica de Canarias, E-38200 La Laguna, Tenerife, Spain\\
    $^2$Departamento de Astrof\'{\i}sica, Facultad de F\'{\i}sica, Universidad de La Laguna, E-38071 La Laguna, Tenerife, Spain}

   \date{Received ; accepted }

   \maketitle
   \markboth{Jordi Cepa and Fernando de Pablos: SFRs, SFEs and IMFs in spiral galaxies}{}

\setlength{\topmargin}{20mm}
\setlength{\baselineskip}{7mm}

\begin{abstract}

A new method of evaluating relative (arm with respect to the interarm disk) star formation rates and relative star formation efficiencies, together with spiral 
arm-amplitudes, as a function of the galactocentric radius, using broad-band 
photometry is derived. The classical method for  obtaining star formation rates 
from H$\alpha$ photometry is discussed, and a new method is derived for diagnosing the 
possible presence of biased star formation due to different initial mass 
functions in the arms and in the interarm disk. As an example, 
these methods are applied to the spiral galaxies NGC 4321 and NGC 4254, 
obtaining their arm amplitudes, relative arm/interarm star formation 
efficiencies, and relative arm/interarm initial mass functions for each arm, 
as a function of the galactocentric radius. Both objects present  
evidence of {\sl massive star formation triggering} in the spiral arms consistent with the spiral density-wave theory, and a {\sl different
initial mass function in the arms from that in the rest of the disk}, in the
sense of favoring a larger fraction of massive stars in the arms. This 
biased star formation is present in the zones of the arms where there is 
triggering of massive star formation, and is then related to, and probably 
caused, by the density-wave system. However, due to this biased star 
formation, the spiral arms of the galaxies studied {\sl do not trigger the 
formation of a larger total mass of stars} (of all spectral types) with 
respect to the interarm disk. 


 \end{abstract}

\setlength{\parskip}{0mm}

\section{Introduction}

Since the first steps by Lindblad (1959) in the search for a theory 
to explain spiral structure in galaxies harmonizing permanent
structures in the disk with its differential rotation---thus avoiding the 
``winding dilemma''---and the formulation of Lin (1968) in what 
would become the spiral density-wave theory, our conceptions about the origin 
of spiral structure and its relation with star formation have evolved, 
although not so dramatically as might be expected. Some points seem 
reasonably established, such as the presence of density waves in the disk (Elmegreen
\& Elmegreen 1984), while others remain controversial. Can the arms propagate through corotation (Patsis et al.1994, and references therein)? Do density waves enhance star 
formation in the arms? Regarding this point, Elmegreen \& Elmegreen 
(1984) found that grand-design spirals (those with continuous, well defined 
arms) have stronger density waves than flocculents (those with patchy and less 
well defined arms), but that the arms of grand-design galaxies were nearly purely 
density wave enhancements. Moreover, from the analysis of several star 
formation tracers, Elmegreen \& Elmegreen (1986) found that the star formation 
rate per unit area is virtually the same in flocculent that in grand-design 
galaxies, concluding that density waves do not significantly enhance star 
formation in the arms with respect to that in the interarm disk. From this and
similar studies, such as that of McCall \& Schmidt (1986), it seemed that 
the mechanism accounting for star formation in the arms would probably be self-propagating star formation as that proposed by Gerola \& Seiden (1978), 
instead of a shock triggered when the material of the disk in its
differential rotation encounters the density wave at supersonic speeds. 
This line of thought was reinforced by the findings of Lubow, Balbus \& Cowie 
(1986), who showed that when taking into account the viscosity of the gas 
and/or gas gravity the gas does not shock. However, Cepa \& Beckman (1990a), 
studied the efficiency of the massive star formation (say the massive star 
formation rate per unit mass of gas) in the arms with respect to that in the 
interarm disk (the so-called {\sl relative massive star formation 
efficiency}) as a function of the galactocentric radius for some grand-design 
spirals and an intermediate-arm type one (M~33), finding that the spiral arms of the grand-design galaxies 
studied were more efficient at forming stars than the interarm disk by a factor 
ranging from 2 up to 15 or more, showing a pattern of peaks and troughs 
coinciding with the position of the resonances and corotation radius predicted by 
the density-wave theory, when a pattern speed is fitted to the curves 
$\Omega_{\rm p}=\Omega\pm\kappa/m$, where $\Omega_{\rm p}$ is the pattern speed, 
$\Omega$ is the rotation curve, $\kappa$ is the epicyclic frequency and $m$ is 
the number of arms. This, and similar results that followed (Cepa \&
Beckman 1990c), led to the conclusion that 
some mechanism is bound to be forming stars in the spiral arms in a different way than 
in the interarm disk and consistently with the spiral density-wave 
predictions. Moreover, this behavior was not observed in the case of 
the intermediate-arm type galaxy studied (M~33), indicating that perhaps
in at least some intermediate-arm type spirals the mechanism forming stars 
in the arms could be a self-propagating one, operating in the arms and in the interarm disk. But, even in this case, some spiral density wave 
has to be present in the disk to allow a self-stocastic star formation 
mechanism to form stable spiral arms, since three-dimensional simulations of 
this mechanism do not form stable spiral arms by themselves (Statler, Comins \& 
Smith 1983).

Some of these results are consistent with those observed by other authors using 
different techniques (Tacconi \& Young 1990 in NGC~6946
; Wilson \& Scoville 
1991 in M~33). Although this could seem to be in contradiction with the studies of 
Elmegreen \& Elmegreen (1986) and Lubow, et al. (1986),it was pointed out by Cepa \& 
Beckman (1990b), on the basis of Elmegreen 
\& Elmegreen's (1986) results, that if the star formation rate is not averaged over the entire
disk and only those parts of the disk where the star 
formation is taking place are taken into account, the star formation rate per unit area {\sl of star 
forming zones} could be as much as 6 times larger in grand-design than in flocculent 
spirals. Also, star formation could be enhanced in the arms by mechanisms 
other than a large-scale shock (Elmegreen 1992, 1994), and hence the expression ``star formation front'' would, in general, be more accurate
than ``shock front''.

However, the studies of relative arm/interarm star formation efficiencies 
mentioned above require spatially resolved measures of the H{\sc i} and CO
over the whole disk of a galaxy, which limits the rate of data adquisition, 
thereby hampering statistically significant results unless large 
amounts of radio observing time are employed. There are also additional 
observational difficulties: the resolution required for this kind of work is 
critical and should be of the order of a star forming region, i.e.,
$\sim$0.5 kpc (Cepa \& Beckman 1990a), otherwise the arm/interarm CO contrast 
is underestimated. This resolution is difficult to achieve 
and slow to map in CO unless interferometric techniques are employed, in 
which case, and given the size of nearby galaxies, a mosaic of observations 
is needed (again slow to map). It could also be difficult to detect interarm 
CO using interferometers. Other critical points are that the conversion factor 
from CO to H$_2$ (Garcia-Burillo \& Gu\'elin 1990), the initial mass function 
(see, for example, Shu, Adams \& Lizano 1987, and references therein), 
and the physical properties of H{\sc ii} regions, could be different in the 
arms from those in the interarm disk, and that the optical measures (to get the star formation rate) could be significantly affected by extinction. All these 
effects may cause severe alterations, either increasing or lowering the measures 
of arm/interarm efficiency ratios by a factor difficult to estimate, depending 
on the galaxy and the galactocentric radius within a given galaxy. 

The methods proposed in the present work allow the measurement of relative 
arm/interarm (or in general from one zone of the disk with respect to 
another) star formation rates and star formation efficiencies as a 
function of the galactocentric radius using optical broad- and 
narrow-band (H$\alpha$ and H$\beta$) imaging observations only. It is possible 
then not only to study these parameters for a larger sample of galaxies in a 
more efficient way, since no radio measurements are needed, but also, from the 
narrow-band data, to correct for extinction and to diagnose possible 
differences between the initial mass function in the arms and the interarm 
zone, thus avoiding most of the uncertainties mentioned in the previous
paragraph. 

The method is described in \S \ 2. Section 3 describes the observations,
data reduction, flux calibration and the procedure followed to apply 
the method to the observed spirals, while the results are discussed in 
\S \  4. Finally, a summary is given in \S \  5. In this paper we
present the method and an example of the application to two grand-design
spirals. In Paper II we will present the results obtained for a larger 
sample of objects of different arm types.

\section{Method}

\subsection{Relative star formation rates}

We consider the total luminosity of a massive star forming zone  (i.e.,
an H{\sc ii} region) in the disk of a galaxy as consisting of the luminosity 
due to the massive star formation plus the luminosity of the underlying older 
disk population. The total luminosity at a given $\lambda$ of a 
massive star forming zone in the interarm disk can then be written as

\begin{equation}
L_\lambda^{\rm IARM}=L_\lambda^{\rm DISK}+L_\lambda^{\rm DSF}
\label{a}
\end{equation}

\noindent where $L_\lambda^{\rm DISK}$ is the luminosity, at a given $\lambda$, of the
underlying older disk, and $L_\lambda^{\rm DSF}$, the luminosity due to the
massive star formation in the interarm disk. Also, the total luminosity 
of a massive star forming zone in the arm is,

\begin{equation}
L_\lambda^{\rm ARM}=A_\star(L_\lambda^{\rm DISK}+L_\lambda^{\rm ASF})
\label{b}
\end{equation}

\noindent where $L_\lambda^{\rm ASF}$ is the luminosity due to the massive star formation, 
and $A_\star$ is the stellar density contrast (in the arm with respect to
the interarm disk). All these quantities depend on the galactocentric radius 
although the dependence is not explicit to simplify the equations. 

The luminosity due to recent star formation can be expressed as a
function of the star formation rate (SFR) $\psi$, the initial mass function
(IMF), $\phi$ and the luminosity, at a given $\lambda$, of the stars of masses
between $m$ and $m+dm$,

\begin{equation}
L_\lambda=\psi \int^{m_{\rm u}}_{m_{\rm l}}\phi (m) L_\lambda(m)\,dm=\psi\,\Phi_\lambda
\label{c}
\end{equation}

Then $\Phi_\lambda$ represent the luminosity per unit mass emitted at 
$\lambda$ of the stars formed, and (\ref{a}) and ({\ref{b}) can be written,

\begin{equation}
L_\lambda^{\rm IARM}=L_\lambda^{\rm DISK}+\psi^{\rm DSF}\,\Phi_\lambda^{\rm DSF}
\label{d}
\end{equation}

\noindent and

\begin{equation}
L_\lambda^{\rm ARM}=A_\star(L_\lambda^{\rm DISK}+\psi^{\rm ASF}\,\Phi_\lambda^{\rm ASF})
\label{e}
\end{equation}

\noindent $\psi^{\rm ASF}$ and $\psi^{\rm DSF}$ are the SFRs in the arm and in the interarm
disk. $\Phi_\lambda$ depends on the IMF and the luminosity of the stars in every range of masses at the given passband $\lambda$, and we will 
consider that the only possible differences between $\Phi_\lambda$ in the 
arm and the interarm region will be caused by differences between the IMFs 
in the arm and the interarm region due to a biased initial mass function.

Using these equations, the relative arm over interarm star formation rate
(RSFR) may be expressed,

\begin{equation}
{\rm RSFR}=\frac{A_\star
\psi^{\rm ASF}}{\psi^{\rm DSF}}=\frac{L_\lambda^{\rm ARM}-A_\star
L_\lambda^{\rm DISK}}{L_\lambda^{\rm IARM}-L_\lambda^{\rm DISK}}\:
\frac{\Phi_\lambda^{\rm DSF}}{\Phi_\lambda^{\rm ASF}}
\label{f}
\end{equation}

\noindent $L_\lambda^{\rm ARM}$, $L_\lambda^{\rm IARM}$ and $L_\lambda^{\rm DISK}$ are
observable quantities that can be obtained performing aperture photometry
on a spiral arm, a massive star forming zone in the interarm disk, and a
zone of the interarm disk with  little or no star formation, respectively.
The passband $\lambda$ has to be sensitive to star formation, i.e., it
has to allow the clear distinction between star forming zones and non-star
forming zones. In this work we will use the Johnson $B$ band, which fulfils
this requirement and is, at the same time, easy to observe (for example,
the Johnson $U$ band requires much longer integration times to achieve the
same signal-to-noise ratio). Also, is reasonably free of emission lines: 
only H$\beta$ lies on the red edge of the filter spectral response curve 
(Allen 1976) at $\sim$50\% of the peak response, or even less, given a
redshifted systemic velocity. 

The stellar density contrast can be estimated from the ratio of arm to
interarm disk luminosities of non-star forming zones in 
a passband less sensitive to star formation, i.e., that approximately
traces stellar density variations. We have used the $I$ band, because it is less 
affected by extinction and is reasonably free of
strong emission lines: [O II] at $\lambda\lambda$ 7325 \AA\ lies on the blue 
edge of the filter at $\sim$65 \% of the peak response (Benn \& Cooper 1987), 
and [S III] at $\lambda\lambda$ 9069 \AA\ falls on the red edge at $\sim$25 \% 
of the peak response (Benn \& Cooper 1987), or even less given a redshifted 
systemic velocity. Then,

\begin{equation}
A_\star=\frac{I^{\rm ARM}}{I^{\rm DISK}}\,K_I
\label{g}
\end{equation}

\noindent where $K_I$ is a factor to correct for the contribution to $I$ band from newly 
formed stars in the arm. From Schweizer (1976) it results that $K_I\sim 0.85$,
in agreement with the observational result of Elmegreen \& Elmegreen (1984).

Finally, using (\ref{g}) and applying (\ref{f}) to Johnson $B$ band, 
(\ref{f}) can be written,

\begin{equation}
{\rm RSFR}=\frac{B^{\rm ARM}I^{\rm DISK}-K_I I^{\rm ARM}B^{\rm DISK}}{I^{\rm DISK}(B^{\rm IARM}-B^{\rm DISK})}
\:\frac{1}{\chi_B}
\label{h}
\end{equation}

In general, $\chi_B=\Phi_B^{\rm ASF}/\Phi_B^{\rm DSF}\ge 1$. If the IMF is the same
over all the disk, $\chi_B=1$. If the star formation is biased towards a
larger fraction of massive stars in the arm with respect to the interarm
disk, $\chi_B>1$. Moreover, in this case the value of $\chi_\lambda$ depends 
on the passband used: the effect of biased star formation (BSF) would be 
more noticeable at shorter wavelengths. 

All the quantities on the right hand side of (\ref{h}) are luminosities
which can be measured directly, except $\chi_B$, which is not an observable 
in the present work. Then,

\begin{equation}
{\rm RSFR}^{\rm meas}={\rm RSFR}\:\chi_B
\label{i}
\end{equation}

\noindent and RSFR$^{\rm meas}$ is the RSFR if no BSF is present, otherwise the
factor $\chi_B$ has to be taken into account. Also, RSFR$^{\rm meas}$
gives an estimate of the relative star formation rate for stars of 
intermediate masses, i.e., the ones that contribute more to the $B$-band 
luminosity.

\subsection{Relative star formation efficiency}

We define the relative (arm with respect to the interarm disk) star
formation efficiency (RSFE) {\sl caused by density waves} as

\begin{equation}
{\rm RSFE}=\frac{{\rm RSFR}}{A_\star}
\label{j}
\end{equation}

\noindent which represents the efficiency of density waves inducing star formation
in the arms. From (\ref{f}): RSFE$=\psi^{\rm ARM}/\psi^{\rm DISK}$. As in the case
of RSFR,

\begin{equation}
{\rm RSFE}^{\rm meas}={\rm RSFE}\:\chi_B
\label{k}
\end{equation}

\noindent Henceforward, the RSFR$^{\rm meas}$ and RSFE$^{\rm meas}$ will be termed 
intermediate-mass star formation rate and efficiency, respectively.

\subsection{Relative massive star formation rates}

The H$\alpha$ luminosity ($L_{{\rm H}\alpha}$) of a star forming zone can be 
related with the number of ionizing photons ($N_{\rm UV}$), assuming case B 
conditions and spherical H{\sc ii} regions via (Osterbrock 1989):

\begin{equation}
L_{{\rm H}\alpha}=h\nu_{{\rm H}\alpha}N_{\rm UV}\epsilon\frac{\alpha^{\rm eff}_{{\rm H}\alpha}}{\alpha_{\rm B}}
\label{l}
\end{equation}

\noindent where $\nu_{{\rm H}\alpha}$ is the frequency of the Balmer $\alpha$ line,
$\epsilon$ measures the efficiency of the gas absorbing photons, and
$\alpha^{\rm eff}_{{\rm H}\alpha}$  and $\alpha_{\rm B}$ depend on the temperature (and
hence on the position of the H{\sc ii} region), and their values are
tabulated for case B recombination (Osterbrock 1989).

By assuming a certain IMF it is possible to relate the H$\alpha$ luminosity 
with the star formation rate: 

\begin{equation}
L_{{\rm H}\alpha} = h\nu_{{\rm H}\alpha} \psi 
\epsilon\frac{\alpha^{\rm eff}_{{\rm H}\alpha}}{\alpha_{\rm B}}
\int_{m_{\rm B}}^{m_{\rm u}}\phi(m)N_{{\rm UV}}(m)\,dm
\label{m}
\end{equation}

\noindent where $m_{\rm B}$ is the lower mass limit for a star to emit photons with a
frequency above the Lyman limit, and $N_{{\rm UV}}(m)$ is the number of 
ionizing photons emitted by a star of mass $m$. This relation will, in
general, depend on the position in the disk (due, for example, to
different metallicities). Also,

\begin{equation}
N_{{\rm UV}}(m)=\int^\infty_{\nu_\circ}\frac{L(m,\nu)}{h\nu}\,d\nu
\label{n}
\end{equation}

\noindent where $L(m,\nu)$ is the luminosity of a star of mass $m$ at frequency
$\nu$, and $\nu_\circ$ is the frequency of the Lyman limit.

Then, if

\begin{equation}
Q = \frac{\epsilon^{\rm IARM}\alpha^{\rm IARM,eff}_{{\rm H}\alpha}\alpha^{\rm ARM}_{\rm B}}
{\epsilon^{\rm ARM}\alpha^{\rm ARM,eff}_{{\rm H}\alpha}\alpha^{\rm IARM}_{\rm B}}
\label{p}
\end{equation}

\noindent and

\begin{equation}
\chi_{{\rm H}\alpha}= \frac{\int_{m_{\rm B}}^{m_{\rm u}}\phi^{\rm ASF}N_{\rm UV}\,dm}
{\int_{m_{\rm B}}^{m_{\rm u}}\phi^{\rm DSF}N_{\rm UV}\,dm}
\label{q}
\end{equation}

\noindent the relative arm/interarm star formation rate inferred from massive stars
can be written, using the same notation as in the previous subsection, 

\begin{equation}
{\rm RSFR}_{{\rm OB}}=\frac{A_\star\psi^{\rm ASF}}{\psi^{\rm DSF}}=
Q\:\frac{{\rm H}\alpha^{\rm ARM}}{{\rm H}\alpha^{\rm IARM}}\:\frac{1}{\chi_{{\rm H}\alpha}}
\label{o}
\end{equation}

\noindent where H$\alpha^{\rm ARM}$ and H$\alpha^{\rm IARM}$ are, respectively, the
H$\alpha$ luminosities of a star forming region in an arm and in the interarm 
disk, respectively, at a similar galactocentric distance $R$ (because
quantities in (\ref{o}) and (\ref{p}) depend on the position in the disk).
 
It is difficult to estimate the ratio $Q$ of (\ref{p}). Let us consider
first the ratio between interarm and arm zone of the term
$\alpha^{\rm eff}_{{\rm H}\alpha}/\alpha_{\rm B}$. This ratio depends on 
the temperature. The densest H{\sc ii} regions could be situated in the arms, 
and densest H{\sc ii} regions are likely to be hotter (McCall, Rybski \& 
Shields 1985). If we assume that in the most extreme case H{\sc ii} regions would have a temperature of 15000 K in the arms and 5000 K in the interarm regions, 
from Osterbrock (1989) it turns out that this ratio takes a value of 
1.1. However, the metallicity of H{\sc ii} regions in the arms might be 
higher, and this effect would tend to lower the temperature (Shields 1990, and 
references therein). Also, the mean temperature of H{\sc ii} regions in the 
arms may well be similar to that in the interarm zone, and then the ratio
would be of order unity. We can then consider that 

\begin{equation}
1.0\le\frac{\alpha^{\rm IARM,eff}_{{\rm H}\alpha}\alpha^{\rm ARM}_{\rm B}}
{\alpha^{\rm ARM,eff}_{{\rm H}\alpha}\alpha^{\rm IARM}_{\rm B}}\le 1.1
\label{r}
\end{equation}

Concerning the ratio between the gas efficiencies in absorbing photons in arm and
interarm zones, McCall et al. (1985) pointed out that the 
H{\sc ii} regions of their sample were ionization bounded,  so that in this case 
$\epsilon$ would be 1.0. Moreover, there is {\it prima facie} 
evidence that the H{\sc ii} regions in the arms and in the interarm zone do 
not differ in their boundary conditions (Cepa \& Beckman 1989; Cepa \& 
Beckman 1990d), thus leading to the conclusion that it is likely that 
$\epsilon^{\rm IARM}/\epsilon^{\rm ARM}=1.0$. However, the regions of the sample of 
McCall et al. were mainly located in the arms, and it may be 
argued that, at least in some cases, H{\sc ii} regions in the interarm zone 
are density bounded while 
H{\sc ii} regions in the arms are ionization bounded. In this case the ratio 
$\epsilon^{\rm IARM}/\epsilon^{\rm ARM}$ would be less than unity. We can do a first-order estimate of this ratio, taking $\epsilon$ as proportional to the 
number of atoms divided by the number of ionizing photons in the H{\sc ii} 
region. The column density of atoms that can be ionized in a cloud is 
proportional to its diameter and density, and the number of 
ionizing photons is proportional to the SFR and IMF. The SFR is itself 
proportional to the product of the gas density and the star formation 
efficiency (SFE). If the SFE is larger in the arms and/or the IMF is 
different in arms and interarm regions, in the sense of favoring the formation of 
more massive stars in the arms, then 
$\epsilon^{\rm IARM}/\epsilon^{\rm ARM}\ge \phi^{\rm IARM}/\phi^{\rm ARM}$, where $\phi$ 
represents the diameter of the H{\sc ii} region. For example, from Cepa \& 
Beckman (1990d), this ratio is 0.75 for NGC 4321, and from Knapen et al. 
(1993) this ratio is 0.72 for NGC 6814. Then, in general,

\begin{equation}
0.7\le\frac{\epsilon^{\rm IARM}}{\epsilon^{\rm ARM}}\le 1.0
\label{s}
\end{equation}

\noindent and finally,

\begin{equation}
0.7 \le Q \le 1.1
\label{t}
\end{equation}

This factor has to be taken into account when evaluating relative arm/interarm 
star formation efficiencies using H$\alpha$ luminosities. In the present
work,

\begin{equation}
{\rm RSFR}_{{\rm OB}}^{\rm meas}={\rm RSFR}_{{\rm OB}}
\:\frac{\chi_{{\rm H}\alpha}}{Q} 
\label{u}
\end{equation}

\noindent and

\begin{equation}
{\rm RSFE}_{{\rm OB}}^{\rm meas}=
\frac{{\rm RSFR}_{{\rm OB}}^{\rm meas}}{A_\star}
\label{v}
\end{equation}

Note that, except for the factor $Q$, ${\rm RSFR}_{{\rm OB}}^{\rm meas}$
and ${\rm RSFE}_{{\rm OB}}^{\rm meas}$ are an estimate of the
relative (arm over interarm disk) star formation rate and efficiency,
respectively, of {\sl massive} stars, and so will be termed henceforward 
when referring to the {\sl measured} quantities.

\subsection{Biased star formation}

The ratio between RSFR and RSFR$_{\rm OB}$ has to be unity. However, this
does not apply to the ratio of the measured RSFRs,

\begin{equation}
\frac{{\rm RSFR}^{\rm meas}}{{\rm RSFR}^{\rm meas}_{{\rm OB}}}=
Q\:\chi
\label{w}
\end{equation}

\noindent where $\chi=\chi_{\rm B}/\chi_{{\rm H}\alpha}$. 

If there is no biased star formation, $\chi=1$. Otherwise, 
$\chi$ would be less than 1.0, because H$\alpha$ samples higher stellar
masses than B. Then the ratio (\ref{w}) could provide an 
observational test for the presence of biased star formation. From (\ref{t}),

\begin{equation}
0.7~\chi \le\frac{{\rm RSFR}^{\rm meas}}{{\rm RSFR}^{\rm meas}_{{\rm OB}}}
\le 1.1~\chi
\label{x}
\end{equation}

\noindent so it is possible to affirm that there is evidence for biased star 
formation in the arms with respect to the interarm disk if ${\rm
RSFR}^{\rm meas}/{\rm RSFR}^{\rm meas}_{{\rm OB}}\le 0.7$.

\section{Observations, data reduction and flux calibration}

\subsection{Observations}

In this paper we present some results derived from the application of the 
method previously presented to two grand-design spiral galaxies, NGC 4321 and 
NGC 4254 (Table 1).

Images were taken for both galaxies in broad ($B$ and $I$) and narrow bands (H$\alpha$ 
and H$\beta$). For narrow-band images we used interference 
filters, including the corresponding continuum filters to the lines H$\alpha$ 
and H$\beta$. To derive the filter transmission in the H$\alpha$ and 
H$\beta$ lines, both the redshift of the galaxy as well as the 
shifts in wavelength in the filter due to temperature changes (Peletier
1994) were taken into account and, when applicable (as in the case of the prime focus of the 
Isaac Newton Telescope), the shift in central wavelength due to the 
non-converging beam (Peletier 1994, see Tables 2 and 3).

The images of NGC 4321 in the $B$, $I$, H$\alpha$ and H$\alpha$
continuum filters were taken in 1994 March with a 1024$\times$1024 pixel Thompson CCD camera 
attached to the cassegrain focus of the 0.8-m IAC-80 telescope at the 
Teide Observatory (Tenerife, Spain), the scale obtained with this configuration being
0.43 arcsec pix$^{-1}$. The H$\beta$ and H$\beta$ continuum images were 
obtained in 1995 June at the Cassegrain focus of the 1.5-m Danish Telescope  
at ESO  (La Silla, Chile), using a 1024$\times$1024 pixel Tektronics CCD, giving 
0.38 arcsec pix$^{-1}$. The images of NGC 4254 were taken in 1995 April with a 
1024$\times$1024 pixel Tektronics CCD camera attached to the prime focus of the 2.5-m 
Isaac Newton Telescope (INT) at the Roque de los Muchachos Observatory (La Palma, Spain). In this case the scale obtained was 0.59 arcsec pix$^{-1}$. Landolt catalogue stars (Landolt 1973) were used for the broad-band calibration. 
For the narrow band we used the spectrophotometric standard stars of Oke \& 
Gunn (1983) and Stone (1977). Atmospheric extinction was evaluated in 
the broad-band filters via the observation of Landolt (1973) standards at 
different airmasses each night or for the INT data using the $V$ extinction 
given by the Carlsberg Automatic Meridian Circle. These
extinctions allow the evaluation of extinctions in other filters assuming a grey
atmosphere and a theoretical extinction curve versus wavelength (King
1985). In Table 4 we present the total integration times, and in Table 5 the 
limiting fluxes and magnitudes attained in the observations. To facilitate the
continuum substraction and to correct for underlying absorption lines
(described in \S \ 3.3 and \S \ 3.4), both the on-line and the continuum images were 
flux calibrated.

\subsection{Data reduction}

Data reduction was performed using the IRAF package. The bias was subtracted
using the overscanned part of the chip. 
Several flatfields were taken for each filter each night and were later 
averaged to obtain a master flatfield for each filter and each night. The sky background was determined for each image by averaging and substracting regions free of 
objects. Afterwards, in the case of the observations 
taken with the IAC-80 or the ESO Danish Telescope, where several 1800 s
exposures were taken for each filter (except for the continuum H$\beta$ at 
the ESO Danish Telescope where the exposure time was 1200 s per image), 
the off-sets among the different images were determined by fitting Gaussians
to several field stars in each frame, the images were aligned to a fraction 
of a pixel and then averaged using a sigma clipping algorithm to get rid of
cosmic ray events. The single exposures obtained with the INT were interactively cleaned
of cosmic rays.

\subsection{Flux calibration and continuum subtraction}

Both on-line and continuum images were flux calibrated using a
procedure similar to that described in Barth {\it et al.} (1994).
The flux calibrated continuum
image was then shifted to a fraction of a pixel to match the on-line image, 
scaled by different factors around 1.00, in steps of 0.05, and
substracted from the on-line image. From the different resulting images,
the one that showed values closer to zero in regions of the interarm disk 
free from H{\sc ii} regions, and no conspicuous negative values in the
bulge zone was selected. The scaling factors found in this way were 1.00 for 
H$\alpha$ images and 0.90 for H$\beta$ images. This procedure for the 
continuum substraction might eliminate diffuse emission coming from the 
interarm disk, but this contribution is irrelevant for the 
present work (see di Serego Alighieri 1987 for an overview of on-line imaging 
and its pitfalls). Figures 1 and 2 show the resulting H$\alpha$ images.

\subsection{Application of the method}

To apply the method outlined here to the data, it is necessary to
evaluate luminosities through three different sets of apertures for each 
galaxy:

\begin{enumerate} 
\item The apertures to measure arm luminosities have been taken
as contiguous, except when there are abundant arm H{\sc ii} regions, where the apertures overlap slightly, and when very few or no arm 
H{\sc ii} regions are apparent, where the apertures undersample the arm. These arm 
apertures are located in the most intense part of the H$\alpha$ image of each arm.
When the H$\alpha$ image presented a gap in the arms, the $B$-band image was used
as a guide to follow the arm in the H$\alpha$ image.
\item The apertures to measure luminosities of star forming (H{\sc ii}) 
regions on the disk have been defined using the H$\alpha$ image.
Zones of doubtful assignation  (to an arm or to the disk), too near to dust
lanes or to the tips of a bar, have been avoided.
\item Finally, aperture positions for zones of the disk without
star formation have been selected avoiding: (i) disk star forming zones
(the H$\alpha$ image has been used for this selection), (ii) emission coming 
from the spiral arms (using the $B$ and $I$ images), and (iii) dust lanes
(using the $B$ and $I$ images). The latter is necessary since it is not possible 
to correct these points for internal extinction, only for Galactic
extinction. Except for the previous stated exceptions, zones equidistant 
from the spiral arms and avoiding bars have been used.
\end{enumerate}

The size of the apertures employed is fixed for each galaxy and  is large enough 
to get isolated H$\alpha$ emission patches of the disk fully inside them. This 
implies diameters of 8$^{\prime\prime}$ for NGC 4321 and NGC 4254 ($\sim$0.8
kpc assuming a distance of 20 Mpc). 

Afterwards, aperture photometry for the whole set of apertures and all the
images (except H$\alpha$ and H$\beta$ and their corresponding continua in the case of disk regions without 
star formation) was performed. The coordinates of the apertures with respect 
to the center of the galaxy, defined as the peak of maximum emission in the 
$I$ band, were then deprojected using the position angles and inclinations 
given in Table 1. 

The luminosities thus obtained were corrected for extinction and
underlying absorption (in the case of H$\beta$) following the procedures
described below.

\subsubsection{Extinction correction}

Since extinction in the disk can be quite large, especially in the spiral arms,
the relative star formation rates must necessarily be corrected for internal 
extinction to be able to apply the method described in \S \ 2 with a 
certain confidence. For instance, extinction can either increase or decrease 
the RSFR, the RSFE and the RSFE$_{\rm OB}$, depending on the zones considered in 
the arms and in the interarm disk, while usually RSFR$_{\rm OB}$, and $A_\star$ 
will be lowered by extinction (assuming that the extinction is larger in the 
arms than in the interarm disk). In the present work, we have used the ratio 
of the Balmer emission lines H$\alpha$ and H$\beta$ to do a first-order 
correction for galactic and extragalactic extinction, using  
the reddening law of Whittet (1992), assuming case B with an electron
density of 100 cm$^{-3}$ and a temperature of 10$^4$ K, from which we obtain

\begin{equation}
A_V = 2.6\, \ln {({F_{\rm H}}_{\alpha} / {F_{\rm H}}_{\beta})_{\rm obs} \over 2.86} 
\end{equation}

\noindent which allows us to derive extinctions for  our photometry. 

This reddening law is not substantially different from that of Schild
(1977). Moreover, recent results (see Davies \& Burstein 1995) show that
it can be applied reliably to external galaxies, at least in the visible
and NIR bands.

However, this method can be applied only to H{\sc ii} regions. 
So, with this method, the luminosities with a superscript 
DISK (\S \ 2) cannot be corrected for internal extinction, only for 
extinction due to our Galaxy and internal extinction for correction to face-on, using $A_B$ from de Vaucouleurs, de
Vaucouleurs \& Corwin (1976). 
In these circumstances, RSFE$^{\rm meas}_{\rm OB}$
is lower, the measured $A_\star$ is larger and if RSFE$^{\rm meas} < 1$ (see
Appendix) RSFR$^{\rm meas}$ and RSFE$^{\rm meas}$ are larger than the same 
fully extinction-corrected quantities. Also, if RSFE$^{\rm meas} < 1$, the 
ratio given by equation (\ref{w}) is larger than if it were fully extinction 
corrected, and then it is still true that if (\ref{w}) is lower than 0.7, 
there is evidence for biased star formation.

\subsubsection{Line-absorption correction}

Apart from extinction, there are several effects that may influence the 
observed fluxes, namely the absorption lines in the calibration 
stars and the underlying stellar absorption of the studied region. The first 
effect is negligible because the equivalent widths of the absorption lines 
are $\simeq$2 \AA\ (McCall et al. 1985)  i.e., small in 
comparison to the filter widths. Hence, the absorbed flux is always 
lower than 4\% of the continuum flux for a filter of 50 \AA\  FWHM or more, 
such as those used in this work. The second effect may be more significant reaching 
reductions in the H$\beta$ Balmer line equivalent widths of 50\% (McCall et al. 1985). This effect leads to an overestimate of the 
absorption. For this correction we used the method given by McCall et al. (1985), which allows us to correct for the underlying absorption 
effect using the equivalent widths of the H$\alpha$ and H$\beta$ lines. 
First, a mask was derived with pixel values equal to unity when the pixels
of the same coordinates in the H$\alpha$ line image had values greater than or equal 
to three times the r.m.s. of the background, and pixel values  otherwise equal to
zero. This mask was multiplied by the continuum images in order
to obtain continuum the contribution of emission-line regions only, not of
regions which are pure continuum. This allows to evaluate equivalent
widths. Since the internal (non-atmospheric) extinction is different in the 
line and in the continuum due to the different central wavelengths of the 
filters used, it is necessary to apply an iterative process to correct for 
underlying absorption and to derive extinctions: initially the H$\alpha$ and 
H$\beta$ equivalent widths uncorrected for extinction are calculated, 
from those we evaluate the
absorption-corrected H$\beta$ fluxes, before deriving the internal
extinction. This internal extinction is used to
correct the H$\alpha$ and H$\beta$ line and continuum fluxes. Using
these first-order extinction-corrected fluxes, we calculate the equivalent widths again, and the H$\beta$ flux is later corrected for 
underlying absorption, which allows the determination of the extinction, etc. This
procedure is repeated until a convergent value is obtained, which takes only
few iterations.

Another effect that might lead to errors in line intensities is 
contamination produced by the [N II] $\lambda$ 6584 {\AA} line. However this 
effect will not affect the arm/interarm ratios, except in the case that 
the variation of the metallicity between the arm and interarm is significant,
since arm/interarm ratios are evaluated at similar galactocentric
distances, thus avoiding the effect of possible metallicity gradients in the 
disk.

Finally, the deprojected, extinction-corrected
luminosities, for each arm, and interarm disk, with and without
star formation, at similar deprojected galactocentric distances, were used 
to apply the formulae of \S \ 2, to determine arm amplitudes,
intermediate-mass star formation rates, massive star formation rates, the
corresponding star formation efficiencies, and look for possible biased 
star formation in spiral arms with respect to the interarm disk as a 
function of radius.

\section{Results}

\subsection{NGC 4321}

NGC 4321 is one of the most conspicuous spiral galaxies in the Virgo cluster. It has two impressive, very well defined spiral arms (of arm class 12
according to Elmegreen \& Elmegreen 1987). The northern arm extends in the 
 east-north-west direction and the southern arm in the  
west-south-east direction.

\subsubsection{Triggered star formation} 

Figures 3 and 4 represent the relative (arm/interarm) intermediate-mass
star formation rate, and the relative massive star formation rate versus 
radius for the southern and northern arm, respectively. 
In both arms the qualitative behavior of both quantities as a function of  radius is noticeably similar for each arm. In the southern arm there is a 
descent below unity of both RSFRs from 105$^{\prime\prime}$ to 
140$^{\prime\prime}$. In the northern arm there is a dip at 
110$^{\prime\prime}$ and a descent below unity at 140$^{\prime\prime}$. 
In these zones, the stellar arm density contrasts are also lower (Fig. 5). 
In the zones where the RSFRs are less than unity, the
interarm disk is forming more stars per unit time than the arm, and can be
identified with parts of the arms with almost no H{\sc ii} regions, as
can be seen in the H$\alpha$ image (Fig. 1). These dips and descents can be 
due to the presence of corotation or to resonances. In the modal theory due 
to Lin and collaborators, oscillations in the amplitude of the wave are 
related to the oscillations of the radial eigenfunctions (see for example Lin 
\& Lau 1979). In the stellar theory for non-linear spirals (Contopoulos
\& Grosb$\o$l 1986, 1988) there is evidence for stellar orbit crowding 
between resonances, and it is possible to find some minima in the arm 
amplitudes placed at different resonances. We assume that corotation is 
located at $\sim$110--120$^{\prime\prime}$, where dips in the RSFRs,  
and stellar density contrasts can be seen in both arms. In this case, 
the dip at 140$^{\prime\prime}$ corresponds to the 4:1 Outer Lindblad 
Resonance (O4:1), while the dip at 60$^{\prime\prime}$ coincides with the 4:1 
Inner Lindblad Resonance, using the rotation curve published by Sempere {\it 
et al.} (1995). The corotation radius determined in this way is consistent 
with that of Sempere {\it et al.} (1995), who found that it is 
located from 85$^{\prime\prime}$ to 115$^{\prime\prime}$, using a test 
based on the H{\sc i} residual velocity field, and that of Elmegreen, 
Elmegreen \& Seiden (1989), at $\sim$120$^{\prime\prime}$, using 
morphological arguments. 

Figure 6 shows the relative intermediate-mass star formation efficiency 
versus radius for both arms of NGC 4321. The arms are less efficient, or as  efficient at forming intermediate-mass stars than the interarm 
(then the reasoning of Appendix A is valid, implying that the RSFE of massive stars is lower, and $A_\star$, RSFR and RSFE of intermediate-mass stars are larger than the same quantities using extinction-corrected disk non-star forming zone luminosities), except beyond 150 arcsec, where measurement uncertainties 
are quite large. In contrast, Fig. 7 shows the relative massive star 
formation efficiency versus radius for both arms of NGC 4321. In some zones,
the southern arm is more efficient at forming massive stars than the interarm 
disk by a factor up to $\sim$2.5--3.5, indicating that density waves are 
triggering massive star formation, a result which was already pointed out 
by Cepa \& Beckman (1990c) using gas density contrasts. However, in the
northern arm, the RSFE$^{\rm meas}_{\rm OB}$ is, in general, less than one. 
Del R\'\i o (1995) and del R\'\i o \& Cepa (1997), using azimuthal 
photometric profiles in Johnson $B$ and $I$ bands, found that the density 
wave is quite inefficient at forming stars in the northern arm, except 
in a localized region outside corotation, which can be identified with 
the region of high RSFE$^{\rm meas}_{\rm OB}$ at $\sim$150$^{\prime\prime}$ 
in Fig. 7. Nevertheless, arm density contrasts are in general larger 
in the northern arm than in the southern arm (Fig. 5), the reason probably being due to the physical conditions of the 
molecular gas. Also, if there is no triggering of massive star 
formation, a linear relation between the RSFR$^{\rm meas}_{\rm OB}$, 
and the arm density contrasts should be expected, and vice versa: triggering 
should show up as a non-linear relation between arm density contrast 
amplitude and relative massive star formation. From Fig. 8 it appears that 
there is no evidence for correlation between the RSFR$^{\rm meas}_{\rm OB}$ 
and the southern arm amplitude, again leading to the presence of some sort of 
triggering mechanism. However, there is a remarkable linear relation between the 
points where no triggering of massive stars is present (with a 
RSFE$^{\rm meas}_{\rm OB}$ 
less than one), which are situated around the corotation radius (Pearson 
correlation coefficient $r_{\rm P}=0.993$ for $n=5$, giving a confidence level of 
99.9 \% for this correlation). In the northern arm, 
avoiding the points being beyond 140$^{\prime\prime}$ (where there is some 
triggering of the star formation), we observe a linear relation between 
RSFR$^{\rm meas}_{\rm OB}$ and the north arm density constrast 
(Fig. 9), with a Pearson correlation coefficient $r_{\rm P}=0.891$ for $n=9$, at 
a 99.8 \% confidence level for this particular correlation. This linear correlation points to the absence of massive star formation triggering in the northern arm, 
since larger arm density contrast corresponds to proportionally larger 
RSFR$^{\rm meas}_{\rm OB}$. However, there is evidence for triggering of the
massive star formation beyond 140$^{\prime\prime}$. This difference in 
behavior between both arms might be due to the interaction of NGC 4321 
with NGC 4322.

\subsubsection{Biased star formation} 

From the previous section, it turns out that the density wave system in NGC
4321 is triggering the formation of massive stars, while intermediate-mass
stars are formed in larger fractions in the interarm disk. This result
suggests that there might be biased star formation in the spiral arms with 
respect to the rest of the disk. As pointed out in 
\S \ 2.4, values of RSFR$^{\rm meas}$/RSFR$^{\rm meas}_{\rm OB}$ 
lower than 0.7 constitute strong evidence for a different 
IMF in the arms with respect to the interarm disk, or, in other words, biased star 
formation (BSF). In Fig. 10 we have plotted this ratio as a 
function of the galactocentric radius for the northern and the southern arms. 
In the northern arm, the ratio of relative measured star formation rates is 
only systematically lower than 0.7 from 130$^{\prime\prime}$ to
150$^{\prime\prime}$, and in 60$^{\prime\prime}$ and 90$^{\prime\prime}$. It would then seem that there is BSF, and that it is mainly situated in the
zone where there is triggered massive star formation. In the southern arm the
ratio is less than 0.7 except in a zone from
100$^{\prime\prime}$ to 130$^{\prime\prime}$. Again, there is BSF, 
mainly in the zone where massive star formation is triggered, and not near
corotation (where no triggering is expected). This is a strong suggestion
that density waves cause a change in the IMF in the arms. This is in some 
agreement with Shu, Lizano \& Adams (1986) who proposed that regions of
higher (lower) star formation efficiency corresponds to regions of
high (low)-mass star formation. A possible reason
might be (Zinnecker 1987) that the giant molecular clouds---located in the 
spiral arms---form massive stars earlier than the standard molecular clouds, the less massive stars not having a chance to form. Also this result is in agreement with that of Scoville, Sanders \& Clemens (1986) who proposed that OB stars form as a result of cloud-cloud collisions, which happen more frequently in the arms. There  
is some other evidence for the presence of BSF in spiral galaxies: Miller 
\& Scalo (1978) conclude that in OB associations in our own Galaxy, which are 
tracers of the spiral arms, only stars with $M \geq 2-5$ $M_{\odot}$ are formed. 
Beyond our Galaxy Rieke et al. (1980) found a 
burst in the center of M82 where they estimate a value of 
$m_l$ $\sim$ 3.5 $M_{\odot}$ for the lower mass limit (see the review
by Shu et al. 1987 for more examples).

\subsection{NGC 4254} 

Iye et al. (1982) found a strong $m=$1 component in the Fourier 
transforms, superimposed on the $m=2$ component, giving the observed $m=3$ 
spiral structure in NGC 4254. Recent studies made by 
Phookun, Vogel \& Mundy (1993) ascribe the cause of the strong $m=1$ component 
to the presence of an extragalactic cloud which merges with the galaxy. 

The optical images clearly show three arms, one of them  appearing
weaker in H$\alpha$ than the others. We have studied the star 
formation in the strongest arms: the northern arm, which extends in the west-north-east direction, and the southern arm, which extends in the east-south-west direction (Fig. 2).

\subsubsection{Triggered star formation} 

In Figs. 11 and 12 we have plotted RSFR$^{\rm meas}$ and
RSFR$^{\rm meas}_{\rm OB}$ versus 
radius for the south and the north arms, respectively. In both arms there is a 
common minimum of both quantities between 85$^{\prime\prime}$ and
95$^{\prime\prime}$, which also appears in the arm amplitudes and
in RSFE$^{\rm meas}_{\rm OB}$, as can be seen in Figs. 13 and 14, 
respectively. From this behavior we should expect that the corotation will 
be between 85$^{\prime\prime}$ and 95$^{\prime\prime}$, a result  
consistent with findings reported by Elmegreen, Elmegreen \& Montenegro (1992), who placed the 
corotation at $\sim$ 87$^{\prime\prime}$. Using this corotation radius, from 
Fig. 14 it turns out that RSFE$^{\rm meas}_{\rm OB}$ in the northern arm is 
greater than unity before and after corotation, and less than unity near or at
corotation, as in the southern arm of NGC 4321. This arm of NGC 4254 contains a region of strong star formation at 115$^{\prime\prime}$ (Fig. 2), which 
has not been included since it is quite possible that it was not triggered 
by the density wave. In fact, the massive star formation efficiency in that 
zone is an order of magnitude greater than in any other zone of star 
formation in this galaxy, and it coincides approximately with the area where 
the extragalactic cloud observed by Phookun et al. (1993) merges 
with the galaxy, this merger being perhaps the cause of the strong star 
formation in the zone. In the southern arm (Fig. 14) there is also some 
triggering of massive star formation mainly after corotation. 

In general, there is no evident correlation between RSFR$^{\rm meas}_{\rm OB}$ and the
stellar density contrast in any arm (Figs. 15 and 16). This fact supports 
the hypothesis that there is triggering of massive star formation in the spiral arms with 
respect to the interarm disk. However, there is a good linear relation for the points in the southern arm having a RSFE$^{\rm meas}_{\rm OB}$ clearly less than one (Fig. 15), i.e., zones where there is no triggering of massive stars (Pearson correlation coefficient $r_{\rm P}=0.994$ for $n=5$, giving a confidence level of 99.9 \% for this correlation). 

As in the case of NGC 4321, the measured RSFEs (Fig. 17) are less 
than or of the order of unity except in some points of the southern arm at
$\sim$30, 40, and 60 arcsec. The density wave system, then, is
triggering massive star formation in the arms, but intermediate-mass stars
are formed preferentially outside the arms, suggesting again the possible
presence of biased star formation in the arms.

\subsubsection{Biased star formation} 

Figure 18 represents the ratio RSFR$^{\rm meas}$/RSFR$^{\rm meas}_{\rm OB}$ 
as a function of the galactocentric radius for the south and the north arm. 
This value is lower than 0.7 in both arms beyond corotation, implying strong 
evidence for BSF after corotation, although a BSF 
before corotation for the southern arm (depending on the values of $Q$) cannot be ruled out. The 
lowest value corresponds to 120$^{\prime\prime}$ in the southern arm, 
correlated with the peak in RSFE$^{\rm meas}_{\rm OB}$ at this distance 
(Fig. 14). Although there seems to be no correspondence between the peak 
in RSFE$^{\rm meas}_{\rm OB}$ at $\sim$80$^{\prime\prime}$ observed in the 
northern arm and a dip in RSFR$^{\rm meas}$/RSFR$^{\rm meas}_{\rm OB}$, the 
observed change of behavior in this ratio before and after corotation,
mainly in the southern arm, suggests that the density wave is responsible for 
the observed BSF.

\section{Summary}

We have developed a procedure for deriving relative (arm/interarm) star
formation rates, arm density contrasts, and relative star formation 
efficiencies as a function of the galactocentric
radius in spiral galaxies, using broad-band (blue and near infrared) and 
continuum-subtracted H$\alpha$ and H$\beta$ imaging. We have also developed a
procedure for assessing whether there is biased star formation in the arms with
respect the rest of the disk, i.e., whether the spiral arms have a different
initial mass function from that of the interarm disk, as a function of radius.

We have applied the previous procedures to the spiral galaxies NGC 4321 and
NGC 4254, finding that density waves greatly influence the star formation and 
initial mass function in the spiral arms.

From the relative star formation rates and arm amplitudes as a function of
the radius, it results that the corotation radius is located at
110--120$^{\prime\prime}$ in NGC 4321 and at 85--95$^{\prime\prime}$ in NGC
4254. 

From the relative intermediate-mass star formation efficiencies and massive 
star formation efficiencies as a function of radius, it is found that 
density waves trigger massive star formation in both galaxies, although not 
with the same strength in the different arms of the same galaxy. However, they 
do not trigger intermediate-mass star formation. Also, triggering is not 
present near the corotation radius where, in the case of NGC 4321, a linear 
relation between relative massive star formation rate and arm amplitude can 
be observed. This linear relation is also apparent in the zones outside 
corotation where the relative massive star formation efficiency is less than 
unity, i.e., where no triggering of massive star formation by the 
density wave is taking place.

Near corotation (and, in the case of NGC 4321, near the Inner and Outer
Lindblad 4:1 Resonances), the interarm disk presents larger star formation
rates and star formation efficiencies than the spiral arms, as if the star
formation were inhibited in the spiral arms near corotation. In fact, a break 
in the arms or a clearly lower number of H{\sc ii} regions in this zone can 
be seen in the H$\alpha$ images.

Intermediate-mass stars form more efficiently
in the interarm disk, and the IMF is different in the arms from that in the disk, 
favoring the formation of a larger fraction of massive stars in the arms. 
This biased star formation seems closely related and probably caused by the 
density-wave system, since it is not present where there is no triggering 
of massive star formation (relative massive star formation efficiencies less 
than unity), in the case of NGC 4321, or it is present after the corotation radius 
and not before, as in NGC 4254.

The observed correlation between biased star formation and higher massive 
star formation efficiency is consistent with the theoretical work of Shu et al. (1986).

Finally, we can conclude that, for the grand-design galaxies studied,
density waves trigger massive star formation, but do not trigger star 
formation of stars of all masses, on the contrary, the interarm disk is
more efficient at forming intermediate-mass stars than the arms, and this is
due to different IMFs in the arms from those in the rest of the disk.

To be able to draw more general conclusions, it is necessary to apply the
method described in this paper to a larger sample of objects, including
intermediate-arm type spirals, where no triggering and consequently no biased
star formation should be expected.

\begin{acknowledgements}

\setlength{\baselineskip}{7mm}

We thank Dr. J. Acosta-Pulido for his useful comments and Dr. A. Manchado for makingthe H$\beta$ on- and
off-images of NGC 4321 available. F. de Pablos is a Resident Astrophysicist at the 
Instituto de Astrof\'\i sica de Canarias.

The IAC-80 Telescope is operated on the island of Tenerife by the
Instituto de Astrof\'\i sica de Canarias at the Spanish Observatorio del 
Teide of the Instituto de Astrof\'\i sica de Canarias. 

The Isaac Newton Telescope is operated on the island of La Palma by the
Royal Greenwich Observatory at the Spanish Observatorio del 
Roque de Los Muchachos of the Instituto de Astrof\'\i sica de Canarias. 

This work has been supported through grant PB94--0433 from the {\em 
Direcci\'on General de Investigaci\'on Cient\'\i fica y T\'ecnica} of the 
Spanish {\em Ministerio de Educaci\'on y Ciencia}.

\end{acknowledgements}

\section*{A. Appendix}

\setcounter{equation}{0}
\renewcommand{\theequation}{A\arabic{equation}}

Differentiating (\ref{h}) considering as variables $I^{\rm DISK}$ and
$B^{\rm DISK}$, and dividing the result by (\ref{h}), we obtain

\begin{equation}
\frac{\Delta {\rm RSFR}}{{\rm RSFR}}=\frac{-\Delta A_\star B^{\rm
DISK}}{B^{\rm ARM}-A_\star B^{DISK}}-\frac{A_\star \Delta B^{\rm
DISK}}{B^{\rm ARM}-A_\star B^{\rm DISK}}+\frac{\Delta B^{\rm DISK}}{B^{\rm
IARM}-B^{\rm DISK}}
\label{aa}
\end{equation}

\noindent where 

\begin{equation}
\frac{\Delta A_\star}{A_\star}=-\frac{\Delta I^{\rm DISK}}{I^{\rm DISK}}
\label{aaa}
\end{equation}

\noindent $\Delta B^{\rm DISK}$ and $\Delta I^{DISK}$ are negative 
quantities that represent the correction to apply to the extinction corrected 
$B^{\rm DISK}$ and $I^{\rm DISK}$, respectively, to obtain the measured
corresponding quantities. Then, $\Delta A_\star > 0$.

However, since extinction is larger at shorter wavelengths, and the colors
of disk non star forming regions are red,
$\vert \Delta I^{\rm DISK}\vert /I^{\rm DISK}<
\vert \Delta B^{\rm DISK}\vert /B^{\rm DISK}$, 
and the second term of (\ref{aa}) is larger than the first. 

If ${\rm RSFE}<1$, then ${\rm RSFR}<A_\star$, and, neglecting the first
term of (\ref{aa}), it results that $\Delta {\rm RSFR}>0$.

Using a similar procedure, it can be demonstrated that if RSFE$< 1$, then
$\Delta {\rm RSFE}> 0$.

\newcommand{\tech}[3]{
                      {#1}        
                      {#2},       
                      {#3\/}     
                      }
%
\newcommand{\journ}[5]{
                      {#1}        
                      {#2},       
                      {#3\/},     
                      {#4},       
                      {#5}        
                      }
%
\newcommand{\inpress}[4]{
                  {#1},        
                  {#2},        
                  {#3\/},      
                  {#4}         
                  }
%
\newcommand{\inprep}[3]{
                  {#1},        
                  {#2},        
                  {#3}         
                  }
%
\newcommand{\book}[4]{
                   {#1}        
                   {#2},        
                   {  #3\/} 
                   {#4}         
                   }
%
\newcommand{\proceed}[6]{
                   {#1},        
                   {#2},        
                in { #3\/}, 
              ed. {#4}        
                   {#5},        
                p. {#6}         
                    }
%
\newcommand{\proceedinpress}[5]{
                   {#1}        
                   {#2},        
                in {#3\/} 
              ed. {#4},        
                   {#5},        
                    }
%
\newcommand{\thesis}[3]{
                   {#1},        
                   {#2},        
                   {  #3\/}, 
                    }

\newpage

\vspace*{-1cm}

\begin{center}
{\bf Figure captions} 
\end{center}

\noindent
{\bf Fig. 1.} H$\alpha$ continuum-subtracted image of NGC 4321. North is
 top and east left. Axes represent offsets in arcseconds from
the nucleus
\\

\noindent 
{\bf Fig. 2.} H$\alpha$ continuum-subtracted image of NGC 4254. North is
 top and east left. Axes represent offsets in arcseconds
from the nucleus. The star formation region marked with an ``X'' sign 
has not been included in the present study
\\

\noindent
{\bf Fig. 3.} Relative massive star formation rate and 
relative intermediate-mass star formation rate as a function of the 
galactocentric radius for the southern arm of NGC 4321
\\

\noindent 
{\bf Fig. 4.} Relative massive star formation rate and 
relative intermediate-mass star formation rate as a function of the 
galactocentric radius for the northern arm of NGC 4321 
\\

\noindent
{\bf Fig. 5.} Arm amplitudes of NGC 4321 as a function of the 
galactocentric radius
\\

\noindent
{\bf Fig. 6.} Relative intermediate-mass star formation efficiency for the
spiral arms of NGC 4321 as a function of the galactocentric radius
\\
 
\noindent
{\bf Fig. 7.} Relative massive star formation efficiency for the
spiral arms of NGC 4321 as a function of the galactocentric radius 
\\

\noindent
{\bf Fig. 8.} Relative massive star formation rate as a 
function of the arm amplitude for the southern arm of NGC 4321 
\\

\noindent
{\bf Fig. 9.} Relative massive star formation rate as a 
function of the arm amplitude for the northern arm of NGC 4321
\\

\noindent 
{\bf Fig. 10.} Ratio of the relative (arm/interarm) intermediate-mass
star formation rate over that of massive stars as a function of the 
galactocentric radius for the northern (upper diagram) and southern (lower 
diagram) arms of NGC 4321. Points below the continuous line (which
indicates the 0.7 value) indicate the presence of biased star formation
\\

\noindent
{\bf Fig. 11.} Relative massive star formation rate and 
relative intermediate-mass star formation rate as a function of the 
galactocentric radius for the southern arm of NGC 4254 
\\

\noindent
{\bf Fig. 12.} Relative massive star formation rate and 
relative intermediate-mass star formation rate as a function of the 
galactocentric radius for the northern arm of NGC 4254
\\

\noindent
{\bf Fig. 13.} Arm amplitudes of NGC 4254 as a function of the 
galactocentric radius
\\

\noindent
{\bf Fig. 14.} Relative massive star formation efficiency for
the spiral arms of NGC 4254 as a function of the galactocentric radius 
\\

\noindent
{\bf Fig. 15.} Relative massive star formation rate as 
a function of the arm amplitude for the southern arm of NGC 4254
\\

\noindent 
{\bf Fig. 16.} Relative massive star formation rate as 
a function of the arm amplitude for the northern arm of NGC 4254
\\

\noindent
{\bf Fig. 17.} Relative intermediate-mass star formation efficiency for the
spiral arms of NGC 4254 as a function of the galactocentric radius
\\
 
\noindent
{\bf Fig. 18.} Ratio of the relative (arm/interarm) intermediate mass
star formation rate over that of massive stars as a function of the 
galactocentric radius for the southern (upper diagram) and northern (lower 
diagram) arms of NGC 4254. Points below the continuous line (which
indicates the 0.7 value) indicate the presence of biased star formation
\\

\newpage

\begin{center}
{\ }
\end{center}

\begin{table}
\caption{Observational parameters of the galaxies observed}
\smallskip
\begin{tabular}{clcrrr}\hline
\noalign{\smallskip}
Galaxy & \multicolumn{1}{c}{Hubble type} & Arm class & 
\multicolumn{1}{c}{P.A.} & \multicolumn{1}{c}{$i$} & 
\multicolumn{1}{c}{$D_{B25}$} \\ 
 & \multicolumn{1}{c}{} & & 
\multicolumn{1}{c}{(deg)} & \multicolumn{1}{c}{(deg)} & 
\multicolumn{1}{c}{(arcsec)}\\
\noalign{\smallskip} 
\hline
\noalign{\smallskip}
NGC4321   &    SAB(s)bc(1)   & 12(2)   & 151(3)   & 31(3)   & 6.9(1)\\
NGC4254 & SA(s)c(1) &  {\ \ 9(2)}  &  58(1)  &27(1)  &   5.4(1)\\
\noalign{\smallskip}
\hline
\end{tabular}

\smallskip
(1) de Vaucouleurs, de Vaucouleurs \& Corwin (1976)\\

(2) Elmegreen \& Elmegreen (1984)\\

(3) Cepa {\it et al.} (1992) 
\end{table}

\newpage

\begin{center}
{\ }
\end{center}

\begin{table}
\caption{Characteristics of the narrow band filters for NGC 4321}
\smallskip
\begin{tabular}{cccc}
\hline
\noalign{\smallskip}
Filter & ${\lambda}_c$  & FWHM & Peak response\\
       &(\AA)			& (\AA)  &    \\
\noalign{\smallskip}
\hline
\noalign{\smallskip} 
H$\alpha$ & 6570 & 95 & 0.63 \\
H$\beta$ & 4897 & 50 & 0.78 \\
H$\alpha C$ & 6260 & 133 & 0.53 \\ 
H$\beta C$ & 4926 & 53 & 0.67 \\
\noalign{\smallskip}
\hline
\end{tabular}
\end{table}

\newpage

\begin{center}
{\ }
\end{center}

\begin{table}
\caption{Characteristics of the narrow band filters for NGC 4254}
\smallskip
\begin{tabular}{cccc}
\hline
\noalign{\smallskip}
Filter & ${\lambda}_c$  & FWHM & Peak response\\
       &(\AA)			& (\AA)  &    \\
\noalign{\smallskip}
\hline 
\noalign{\smallskip}
H$\alpha$ & 6607 & 53 & 0.54 \\ 
H$\beta$ & 4863 & 97 & 0.68 \\ 
H$\alpha C$ &  6565 & 45 & 0.56 \\
H$\beta C$ & 4540 & 114 & 0.53 \\
\noalign{\smallskip}
\hline
\end{tabular}
\end{table}

\newpage

\begin{center}
{\ }
\end{center}

\begin{table}
\caption{Total exposure times in seconds per filter for NGC 4321 and NGC 4254. 
The aperture of the telescope used is given in brackets}
\smallskip
\begin{tabular}{crr}
\hline
\noalign{\smallskip}
Filter & \multicolumn{1}{c}{NGC 4321} & \multicolumn{1}{c}{NGC 4254} \\
\noalign{\smallskip}
\hline
\noalign{\smallskip}
H$\alpha$ & 7200 (0.8 m) & 1800 (2.5 m) \\
H$\alpha C$ & 5400 (0.8 m) & 1800 (2.5 m) \\
H$\beta$ & 5400 (1.5 m) & 1800 (2.5 m) \\
H$\beta C$ & 3600 (1.5 m) & 1800 (2.5 m) \\
$B$ & 7200 (0.8 m) & 1200 (2.5 m) \\
$I$ & 7200 (0.8 m) & 600 (2.5 m) \\
\noalign{\smallskip}
\hline
\end{tabular}
\end{table}

\newpage

\begin{center}
{\ }
\end{center}

\begin{table}
\caption{Limiting fluxes and magnitudes for NGC 4321 and NGC 4254, at three
times the r.m.s. of the sky background}
\smallskip
\begin{tabular}{cll}
\hline
\noalign{\smallskip}
Filter & \multicolumn{1}{c}{NGC 4321} & \multicolumn{1}{c}{NGC 4254} \\
\noalign{\smallskip}
\hline
\noalign{\smallskip}
H$\alpha$ & $1.1\cdot 10^{-16}$ erg cm$^{-2}$ s$^{-1}$ & 
$3.4\cdot 10^{-17}$ erg cm$^{-2}$ s$^{-1}$ \\
H$\beta$ & $5.1\cdot 10^{-18}$ erg cm$^{-2}$ s$^{-1}$ & 
$6.0\cdot 10^{-17}$ erg cm$^{-2}$ s$^{-1}$ \\
$B$ & 23.34 mag arcsec$^{-2}$ & 25.40 mag arcsec$^{-2}$   \\
$I$ & 22.18 mag arcsec$^{-2}$ & 23.10 mag arcsec$^{-2}$\\
\noalign{\smallskip}
\hline
\end{tabular}
\end{table}

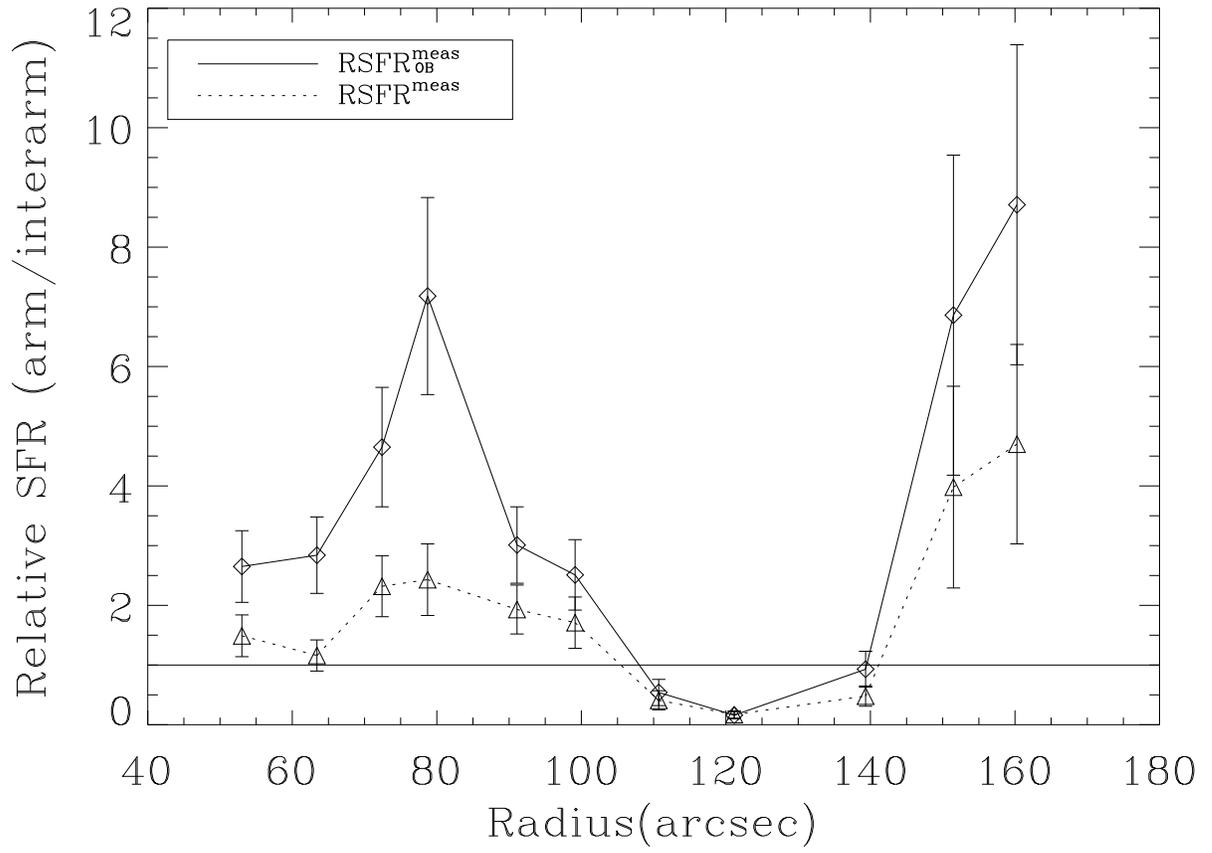
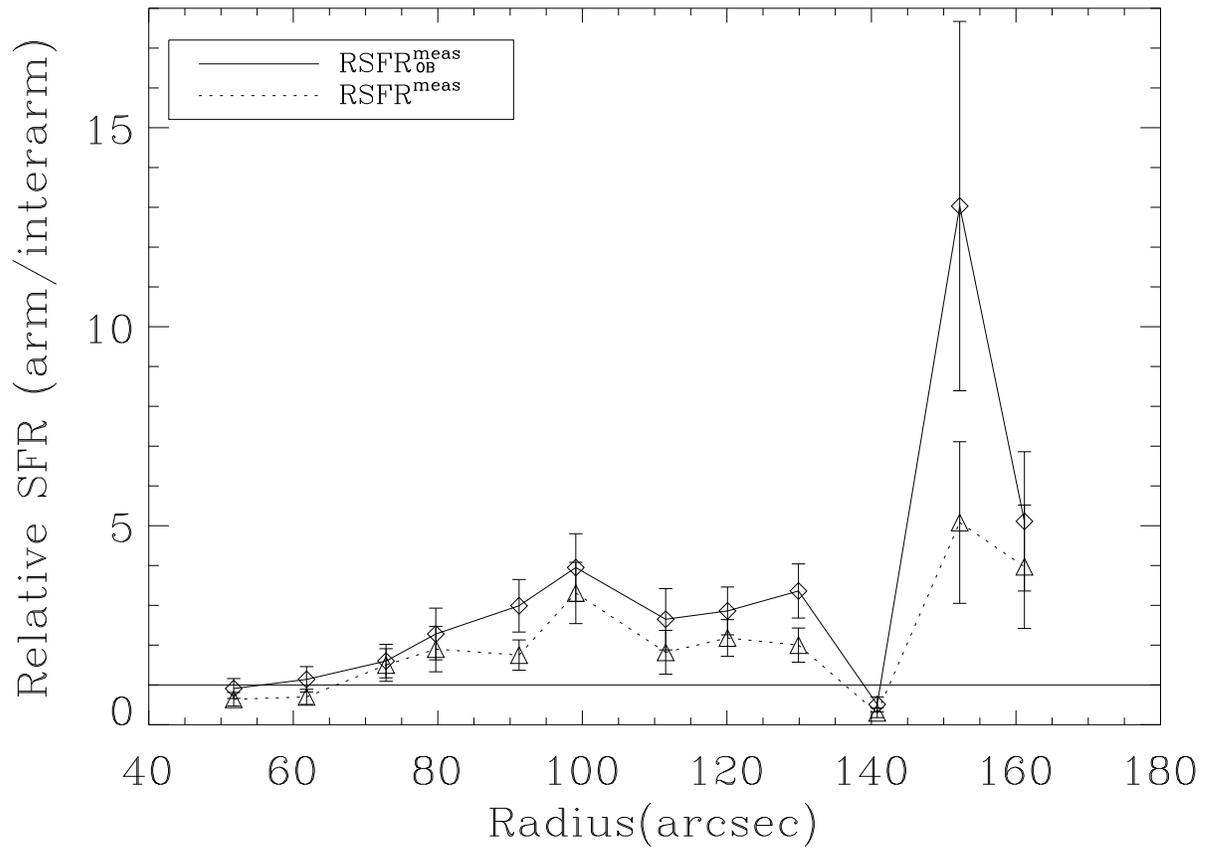
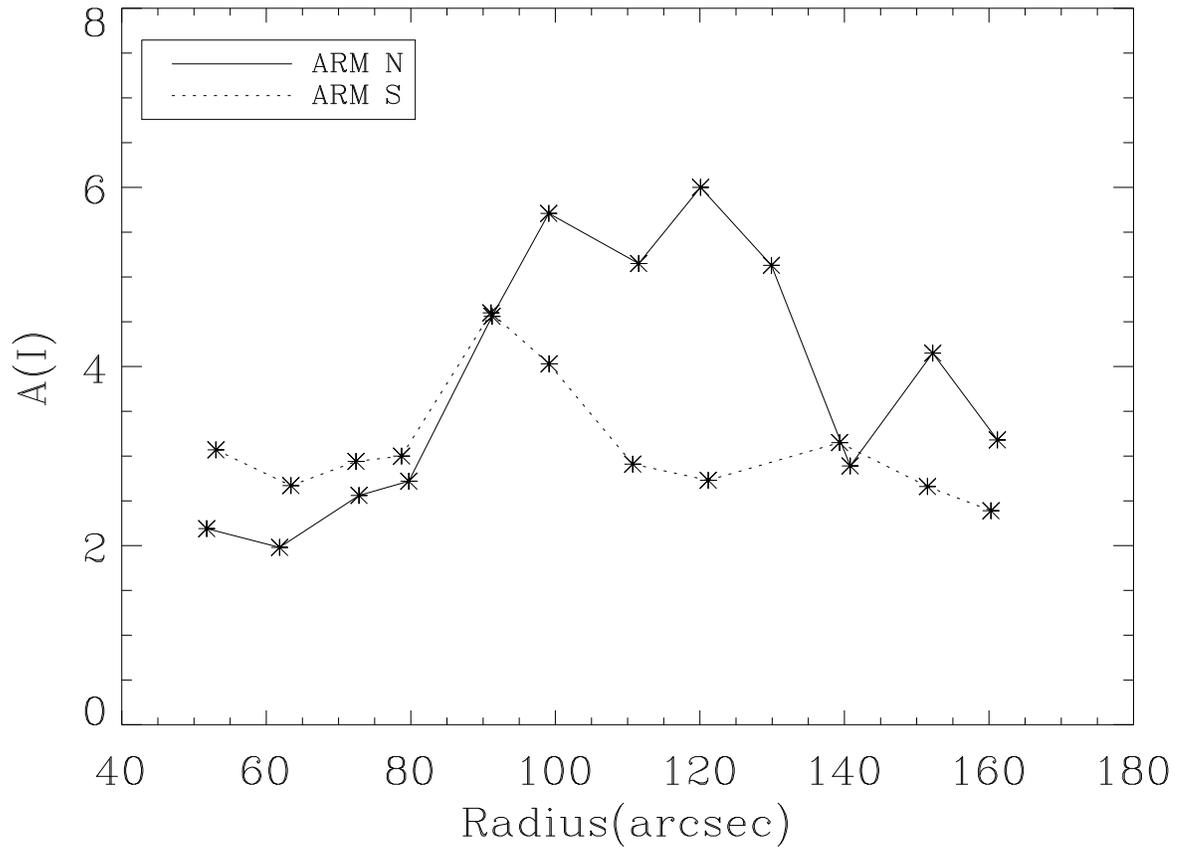
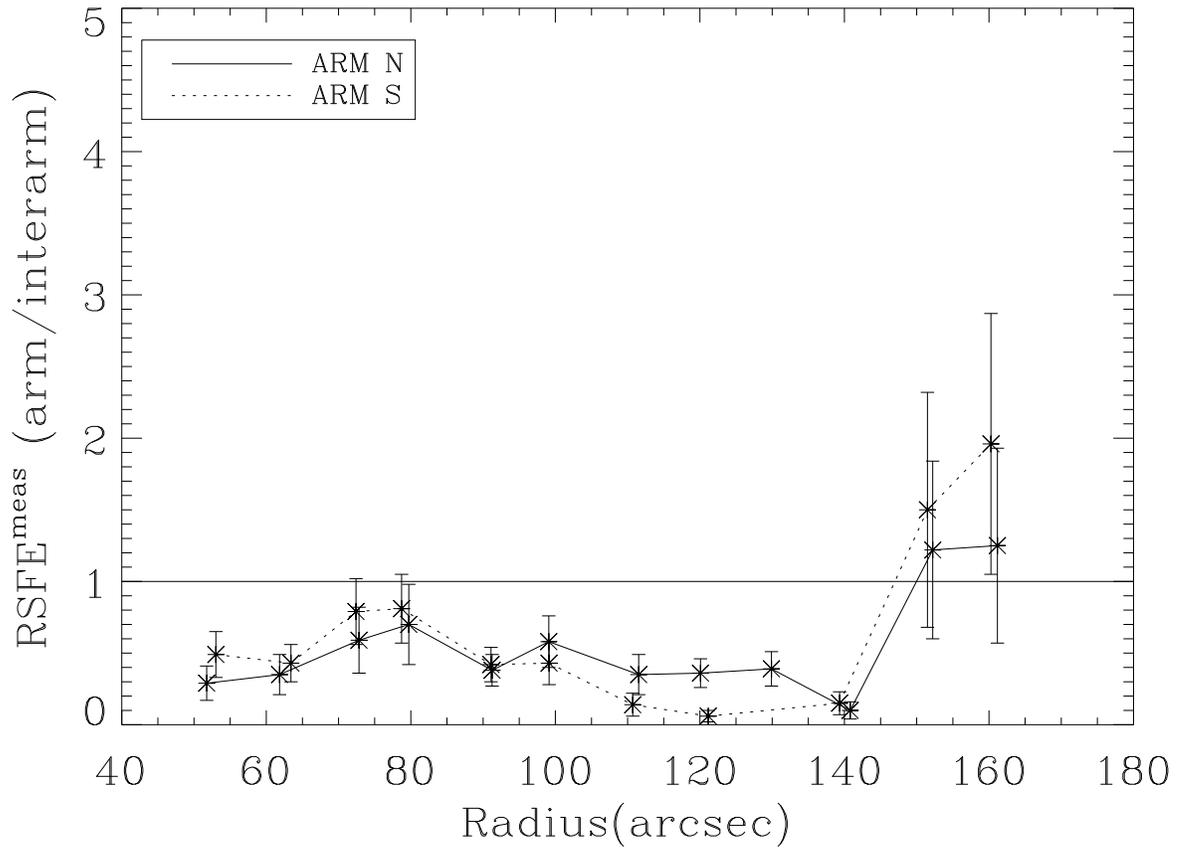
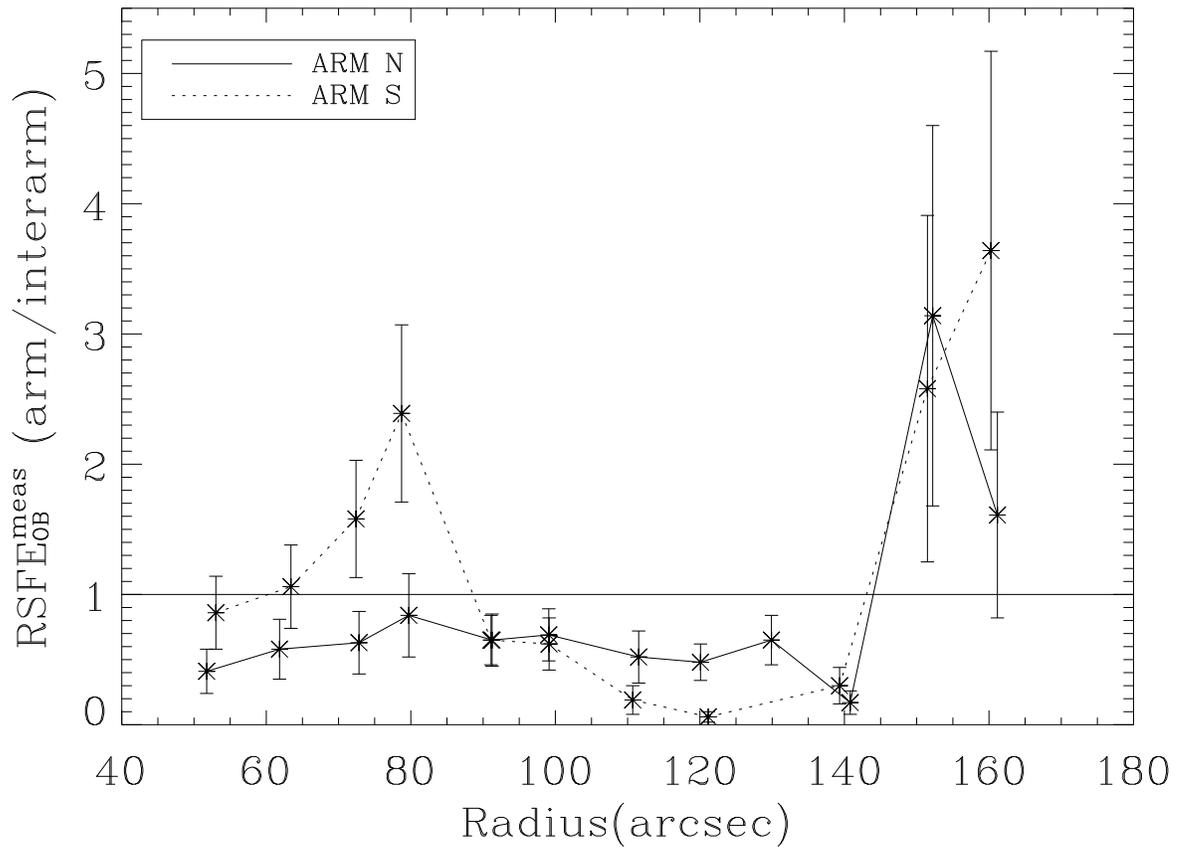
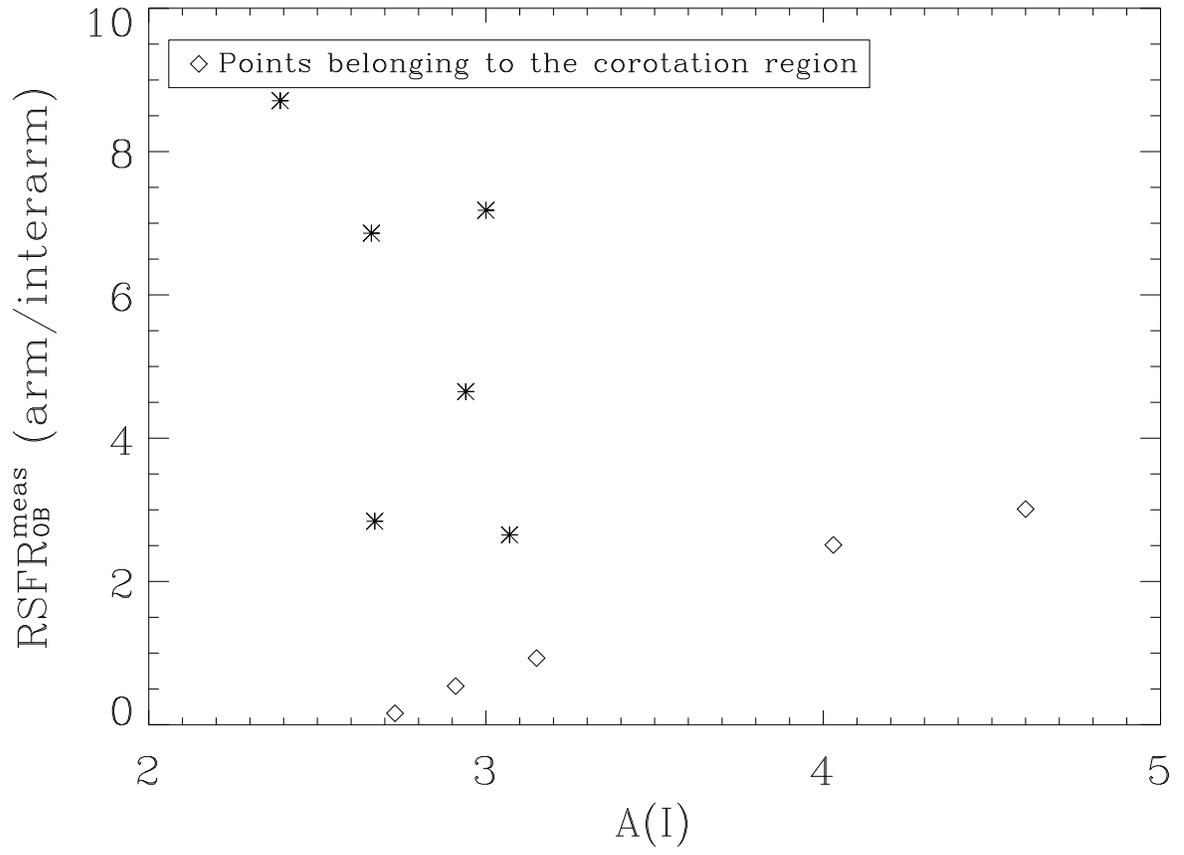
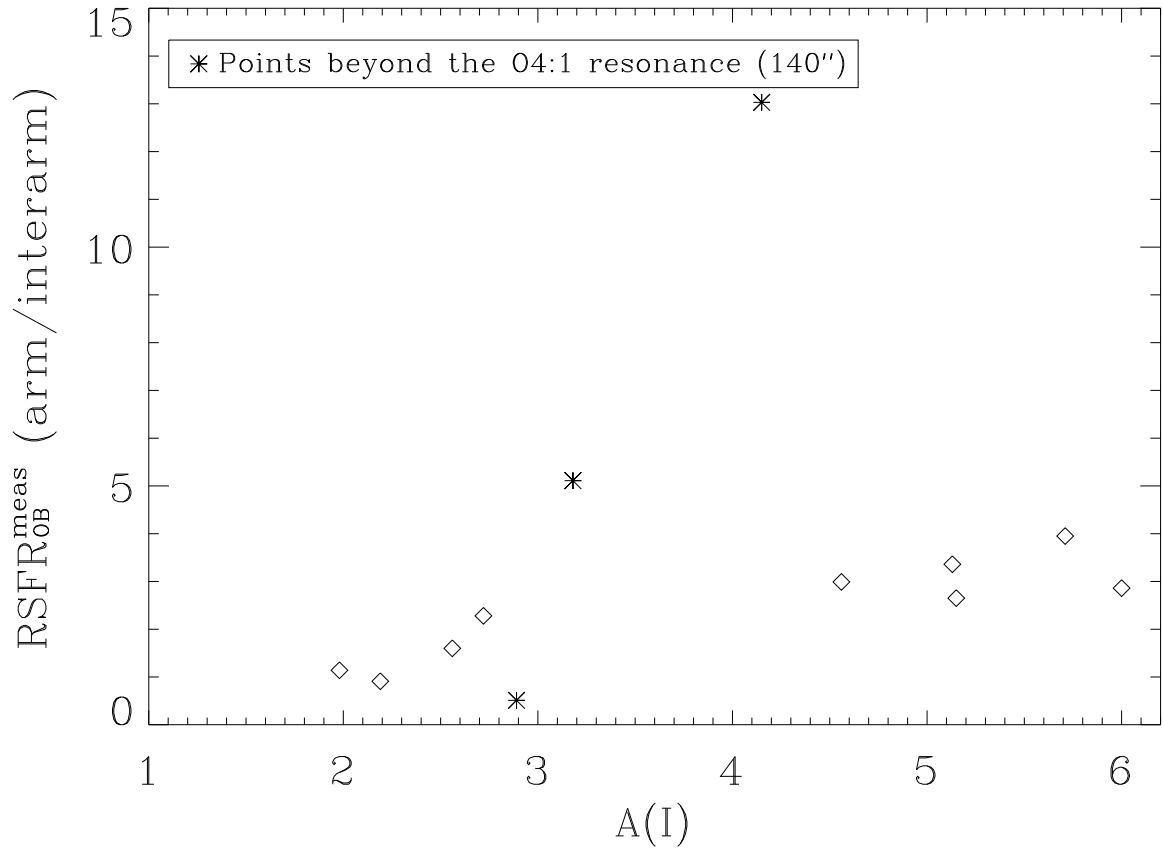
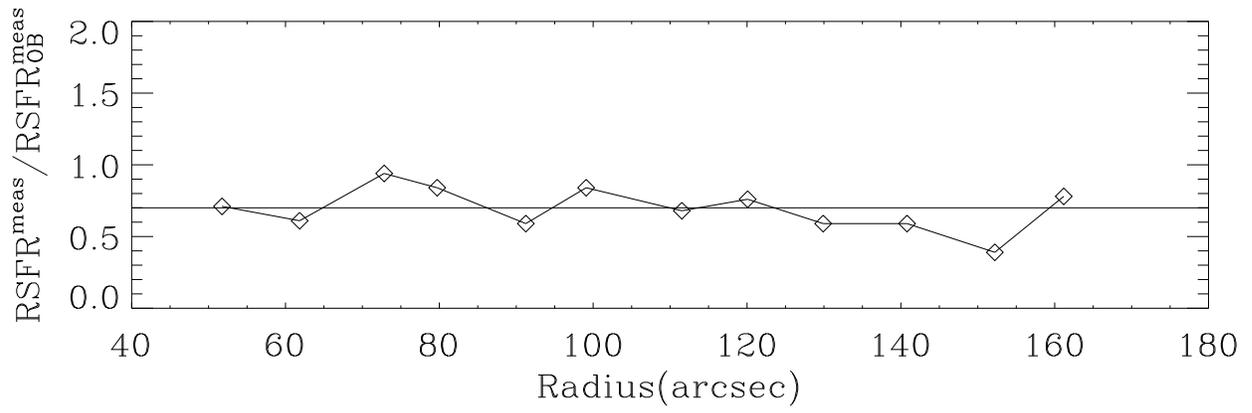
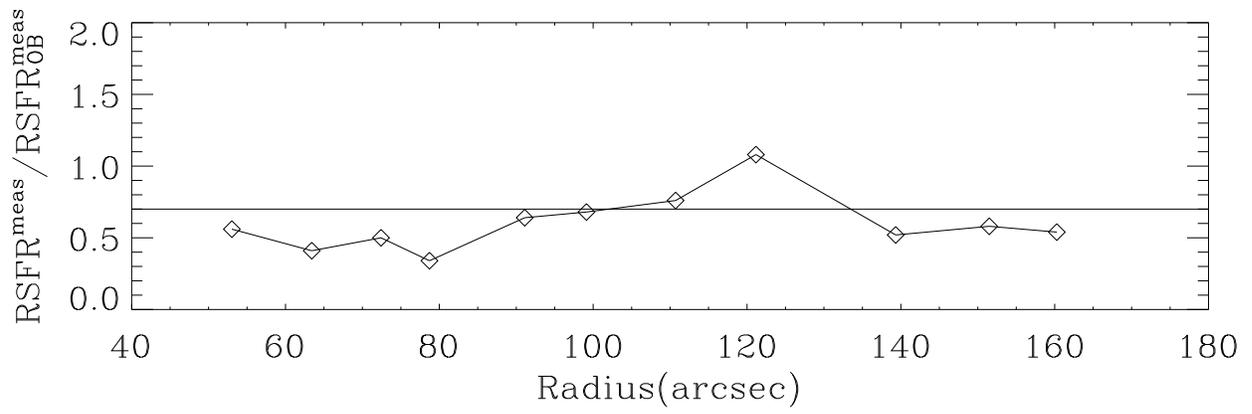
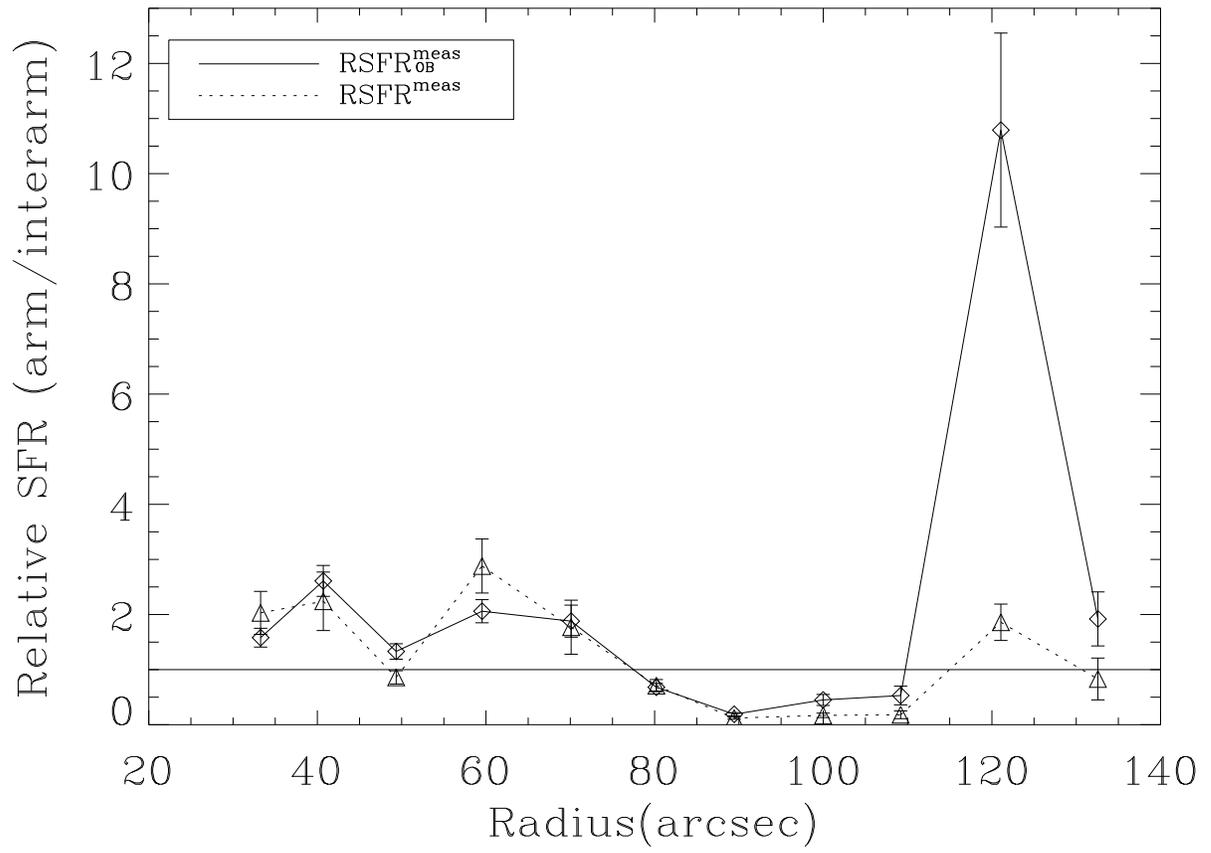
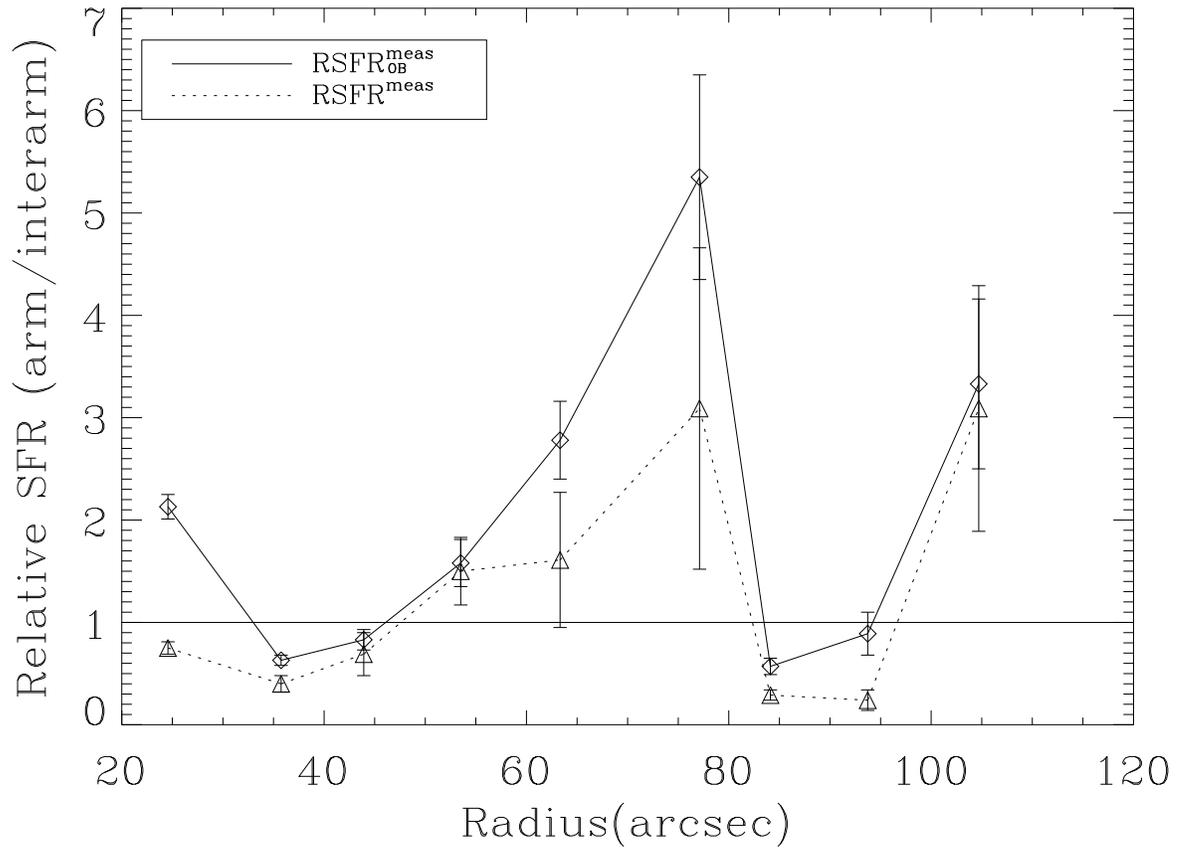
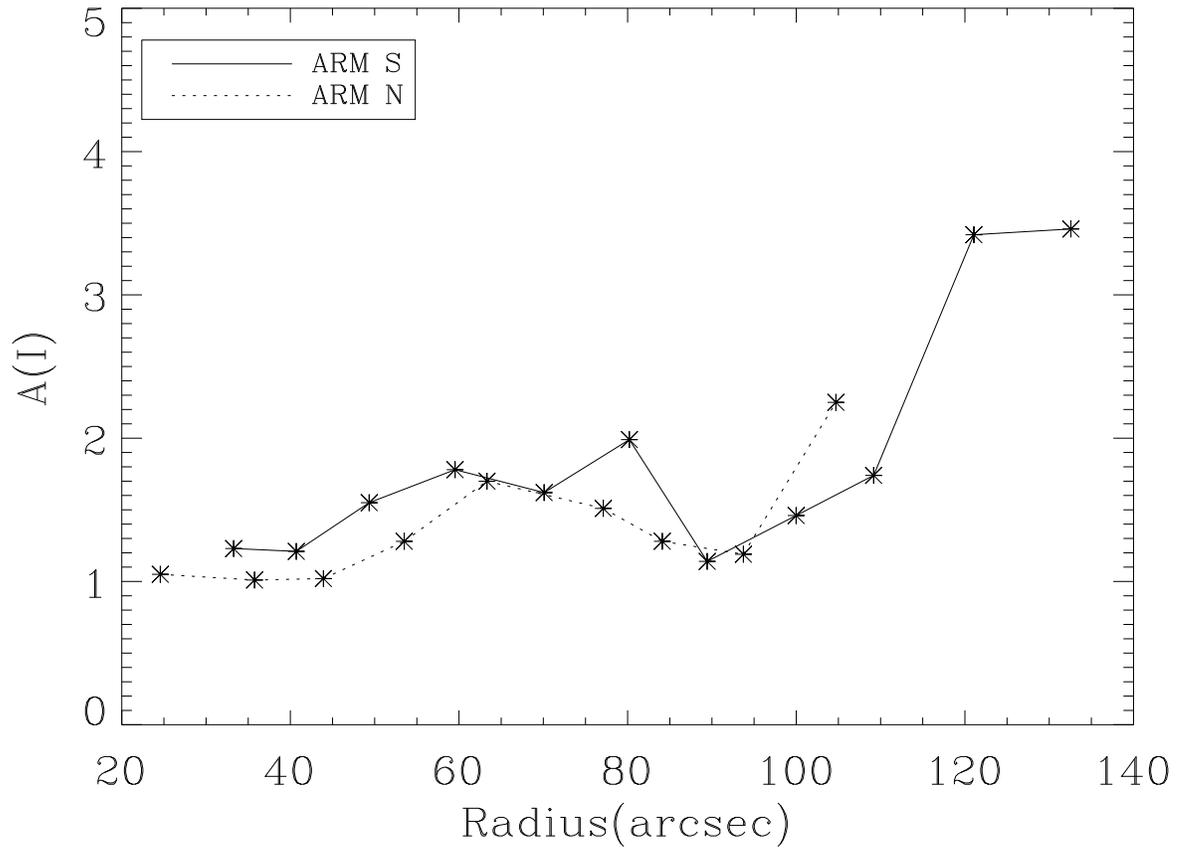
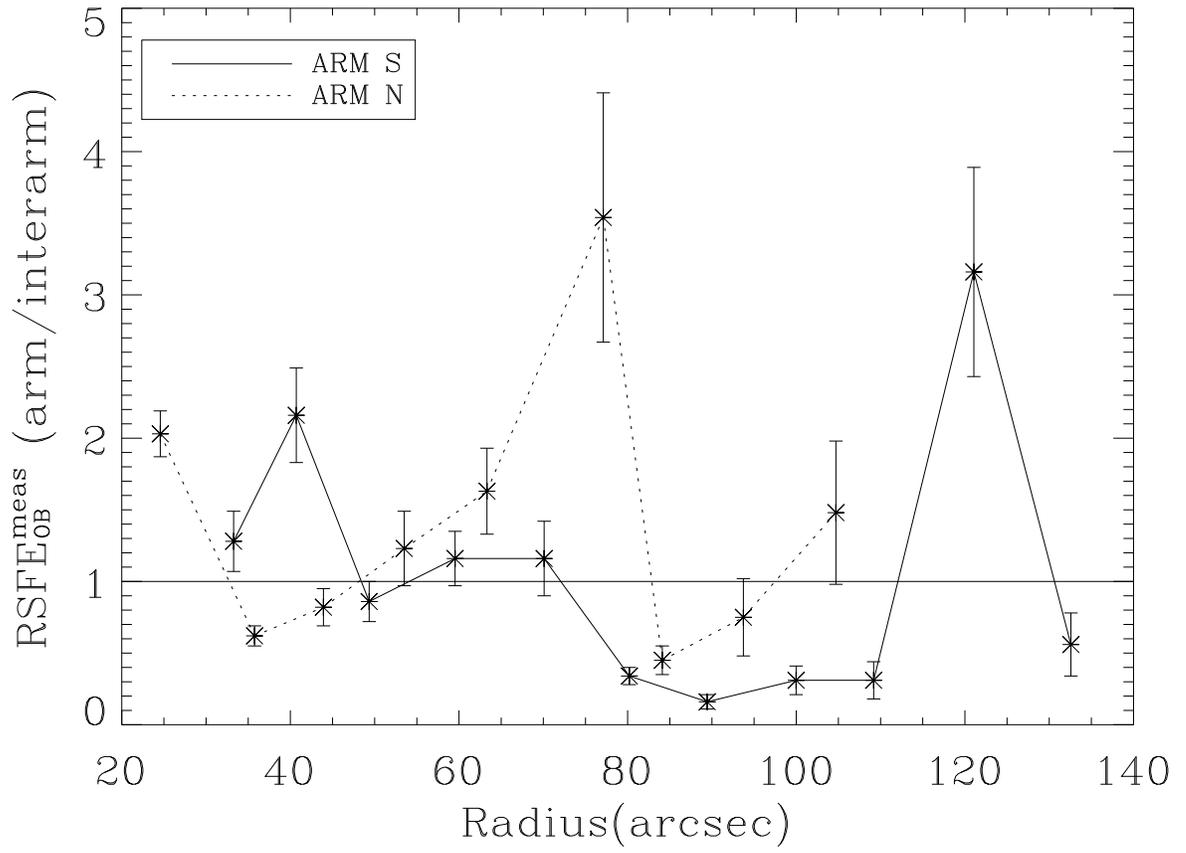
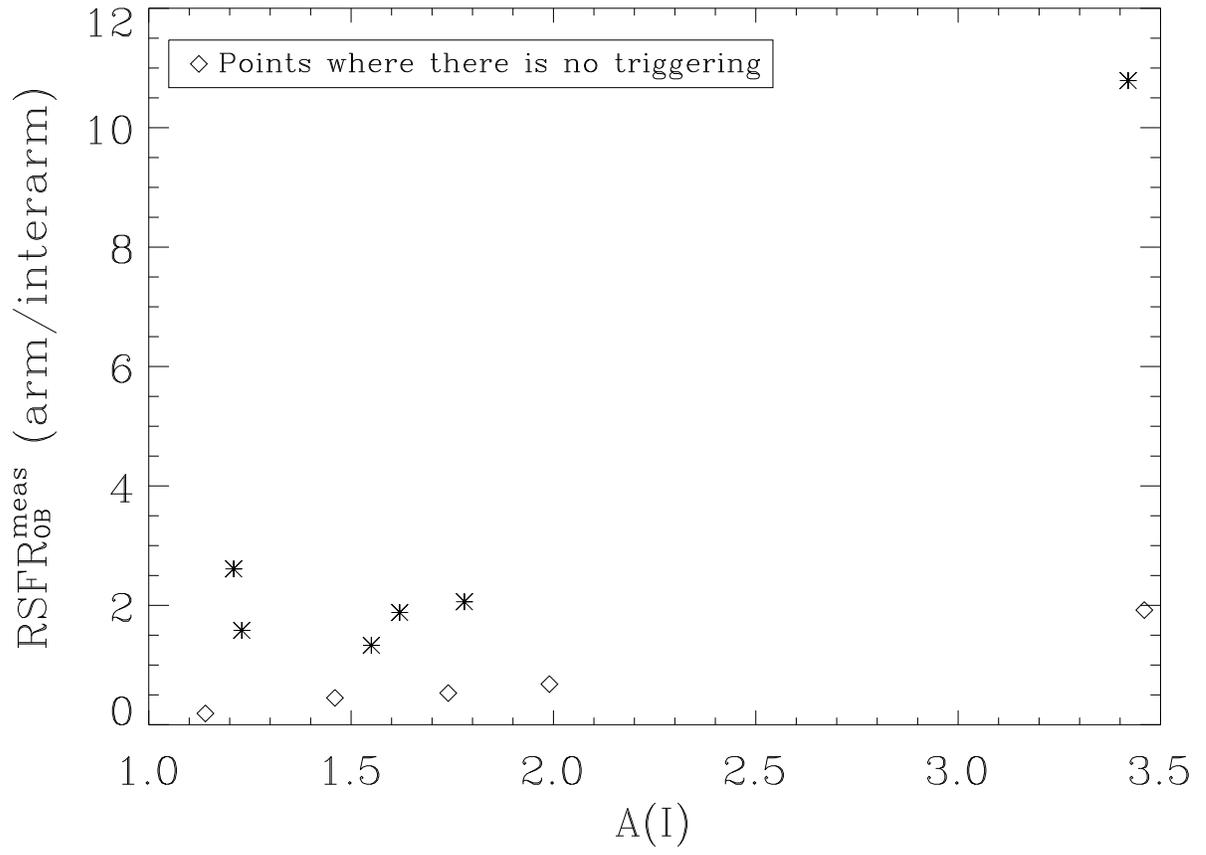
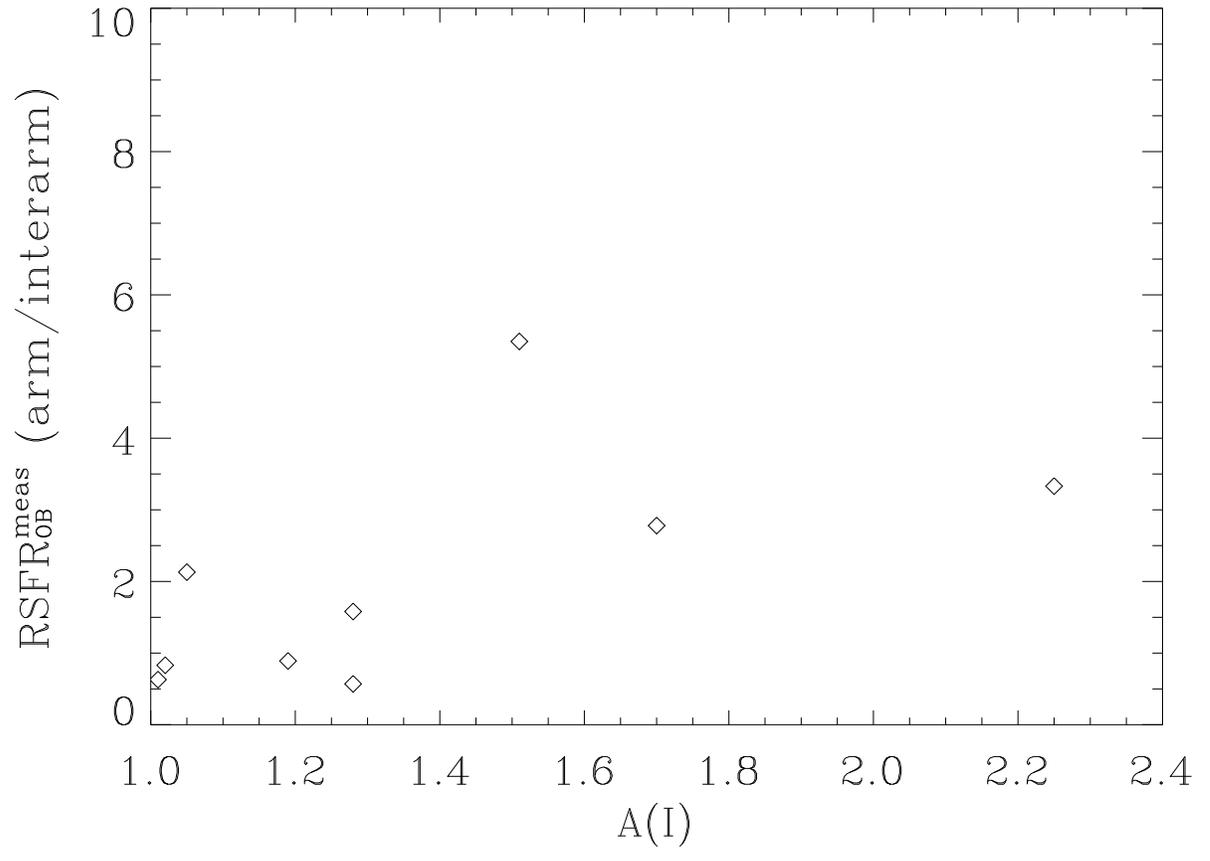
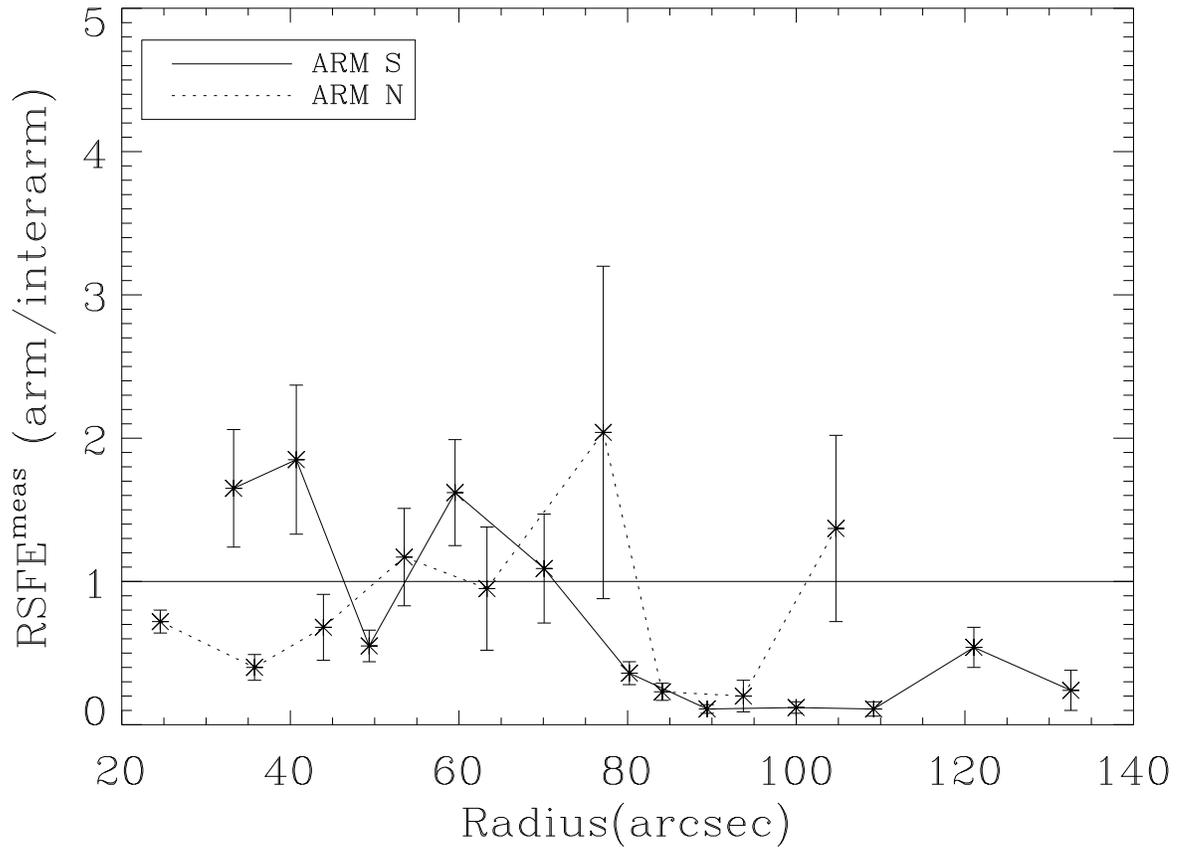
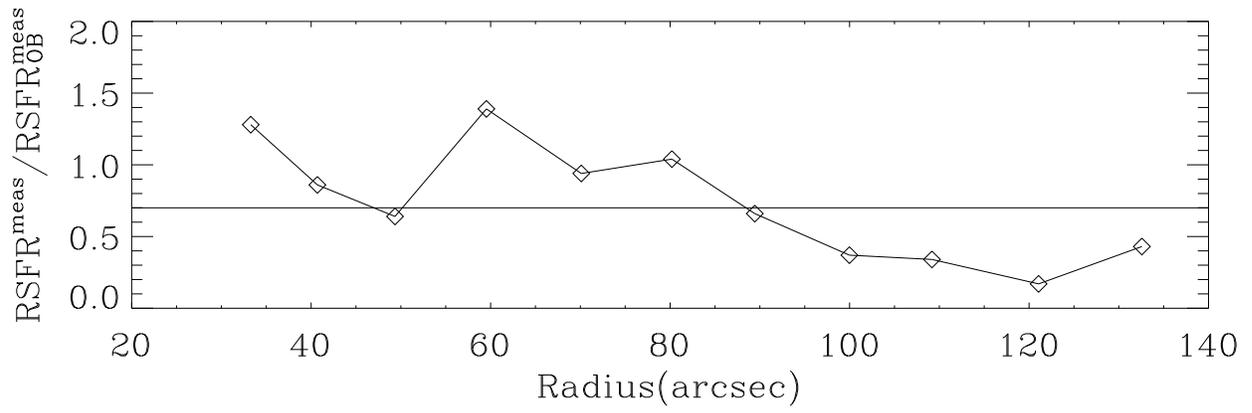
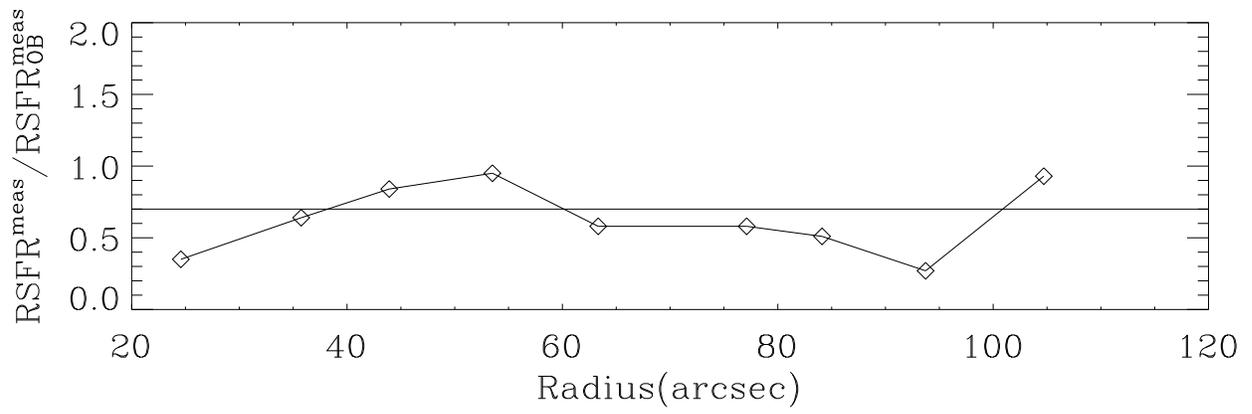
\end{document}